\documentclass[]{spie}  

\usepackage{amsmath,amsfonts,amssymb}
\usepackage{graphicx}
\usepackage[colorlinks=true, allcolors=blue]{hyperref}
\usepackage{array} 
\usepackage{siunitx}
\graphicspath{ {./images/} }

\title{Investigating Mass Reduction Capabilities of Additive Manufacturing through the Re-Design of a Space-Based Mirror}

\author[a]{Rhys Tuck}
\author[a]{Younes Chahid}
\author[a]{Greg Lister}
\author[a]{Katherine Morris}
\author[a]{James Carruthers}
\author[b]{Mat Beardsley}
\author[b]{Michael Harris}
\author[b]{Michal Matukiewicz}
\author[c]{Simon Alcock}
\author[c]{Ioana Theodora Nistea}
\author[a]{Carolyn Atkins}

\affil[a]{UK Astronomy and Technology Centre, Royal Observatory, Edinburgh, EH9 3HJ, UK}
\affil[b]{RAL Space, Harwell Science and Innovation Campus, OX11 0QX, UK}
\affil[c]{Diamond Light Source, Harwell Science and Innovation Campus, OX11 0QX, UK}

\authorinfo{Further author information: \\ E-mail: carolyn.atkins@stfc.ac.uk \\ E-mail: r\_tuck@outlook.com}

\begin{document} 
\maketitle

\begin{abstract}
Additive manufacture (AM) involves creating a part layer by layer and is a rapidly evolving 
manufacturing process. It has multiple strengths that apply to space-based optics, such as 
the ability to consolidate multiple parts into one, reducing the number of interfaces. The process also 
allows for greater mass reduction, making parts more cost-effective to launch, achieved by optimising 
the shape for intended use or creating intricate geometries like lattices. However, previous studies 
have highlighted issues associated with the AM process. For example, when trying to achieve high-precision optical surfaces on AM parts, the latticing on the underside of mirrors can provide 
insufficient support during machining, resulting in the quilting effect. 
This paper builds on previous work and explores such challenges further. This will be implemented by 
investigating ways to apply AM to a deployable mirror from a CubeSat project called A-DOT. The 
reflective surface has a spherical radius of curvature of 682 mm and approximate external 
dimensions of 106 x 83 mm. The aim is to produce two mirrors that will take full advantage of AM 
design benefits and account for the challenges in printing and machining a near-net shape. The 
designs will have reduced mass by using selected internal lattice designs and topology-optimised 
connection points, resulting in two mirrors with mass reduction targets of 50\% and 70\%. Once 
printed in aluminium using laser powder bed fusion, the reflective surface will be created using 
single point diamond turning. Finally, an evaluation of the dimensional accuracy will be conducted, 
using interferometry, to quantify the performance of the reflective surface.  
\end{abstract}

\keywords{Additive Manufacturing, Lightweight Mirrors, Surface Finish, Space-Based Optics, Topology Optimisation, Laser Powder Bed Fusion, CubeSat}

\section{INTRODUCTION}
\label{sec:introduction}  

\begin{figure}
    \centering
    \includegraphics[width = 0.95\textwidth]{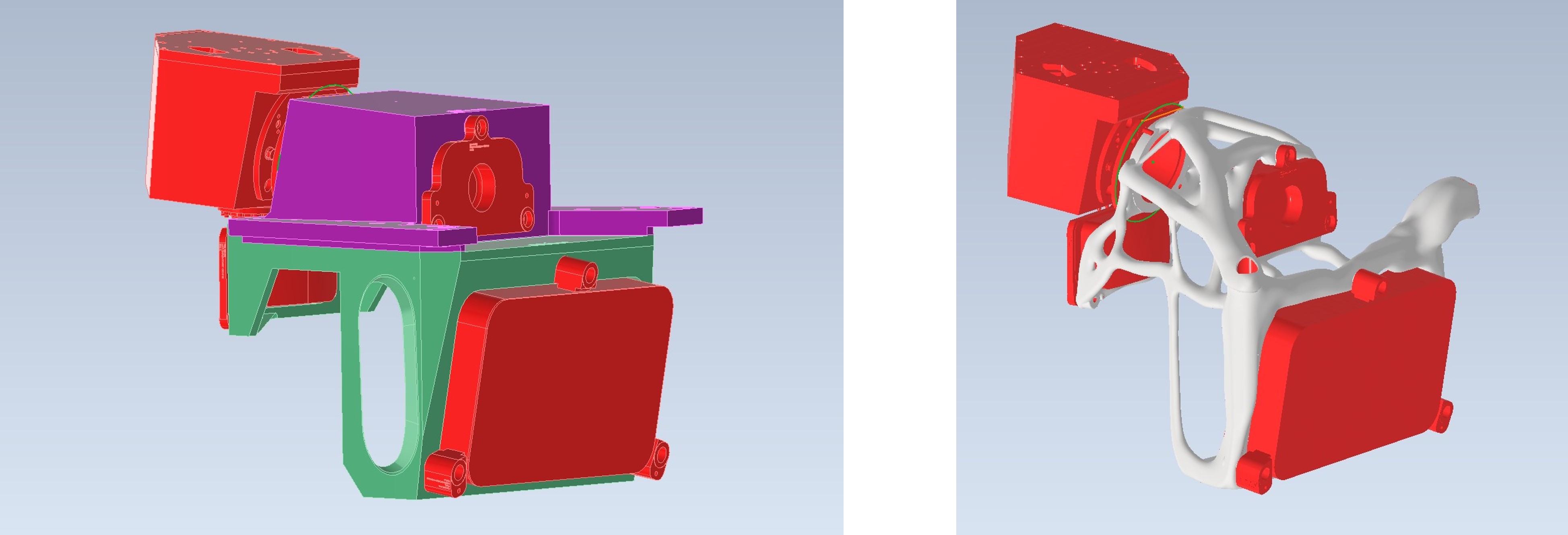}
    \caption{Topology optimised frame to hold mirrors (red), adapted from \textit{Wells, J., et al. (2023)} \cite{JamesPaper}: \textit{Left} original frame, \textit{Right} topology optimised design}
    \label{fig:1}
\end{figure}

Additive Manufacturing (AM) is still a relatively new form of manufacture when compared with the more established techniques of subtractive (turning/milling), fabricative (gluing/welding) and formative (casting/moulding), especially when applied to the manufacture of lightweight mirrors. In the AM process, a geometry is constructed a layer at a time, with each layer fusing to the one below it, until the complete shape is formed. This process enables the creation of more complex geometries and features than conventional manufacturing methods allow. One of the techniques that makes use of this is topology optimisation. This process analyses stress to identify high and low-stress areas within a design volume. Material in low-stress regions is removed, often resulting in organic-styled geometries as seen in Figure~\ref{fig:1}. The primary benefit of this design technique is a greatly improved mass-to-strength ratio through more economical use of material. Features designed like this or others such as intricate lattices, enable different designs with benefits that were not easily manufactured before. Another advantage of AM is the ability to take multiple parts and fasteners for an assembly and consolidate them down into one part, simplifying both the assembly and function.

AM has lots of applications and the aerospace\cite{AeroSpaceBackground} and medical fields\cite{MedicalAMReview}were early adopters. The advantages mentioned above also lend themselves very well to space optics. The cost associated with the launch of a satellite is correlated with the the mass of the satellite, so reducing the mass of components is very useful in lowering the overall cost for satellite production. Reducing the number of parts is another attribute that is advantageous in space applications. Designs with fewer parts not only reduce assembly time but also eliminate potential failure points that could cause irreversible issues in operation. \textit{Horvath, N. et al (2020)}\cite{Horvath:21} is a good example of the potential benefits of AM, showing a comparison between a traditionally designed lightweight mirror and a mirror lightweighted by designing it for AM. While both design methodologies had challenges, the AM design emerged as the more promising.

These benefits have thus prompted further research into the feasibility of mirror fabrication through the AM process. \textit{Mici, J. et al (2015)} \cite{JoniMici} was an early paper that explored the idea of light-weighting a mirror through AM. The mirror in this case was a reflective optic used in high-energy laser applications. There was a lot of crossover with space optics, particularly in the analysis of lattices used to lightweight the structure.  The paper by \textit{Atkins, C., et al. (2017)} \cite{CAtkins2017} demonstrated an early practical attempt to print and polish a mirror with a space application in mind, printing with three different methods and materials to obtain metrology data on a polished optical surface. Topology optimisation was then used on promising materials to better utilise the lightweight capabilities of the technology. After polishing, the RMS form error and surface roughness were measured as \SI{38.6}{\nm} and \SI{4.20}{\nm} respectively for an aluminium print coated in nickel phosphorous. This was the mirror that produced the best optical surface from this study and achieved a weight reduction (WR) of about 33\% from solid aluminium. The WR was ultimately limited by the nickel phosphorous layer that was coated on the top surface. This layer was however necessary to provide a harder and smoother substrate to polish, as the uncoated aluminium surface produced poorer which were largely due to the softer nature of aluminium as a material and the porosity present.

The porosity in AM printed aluminium poses a significant issue when trying to achieve a good optical surface on a mirror. Most AM aluminium mirrors are printed using a form of Metal Laser Sintering called Laser-Powder Bed Fusion (L-PBF)\cite{L-PBFOverview}. The process consists of selectively melting a thin layer of powder with a laser to form solid metal, another layer of powder is then coated on top and this layer is melted to the previous layer, this is repeated until a new solid metal part can be extracted from the surrounding, un-melted, powder. Porosity therefore occurs in two ways, when the laser fails to melt pockets of material or when it over-melts regions, causing gas to form while the material solidifies. Research by \textit{Snell, R., et al. (2022)} \cite{R.Snell-2022} and \textit{Atkins, C., et al. (2019a)} \cite{C.Atkins-2019-01} investigates this issue further. The extent of the issue was quantified by using ``X-ray computed tomography measurements" to ``highlight the location and density of pores within the mirror substrate"\cite{C.Atkins-2019-01}. Efforts were then made to understand ``the relationships laser parameters, laser paths and build orientations have with the porosity"\cite{R.Snell-2022} and possible solutions, such as ``hot isostatic press"\cite{R.Snell-2022}

The issue of print-through or the `quilting effect' is also discussed in literature\cite{Carolyn2019/2,Marcell,Horvath:21,ZhangKai}. It is a common challenge when polishing a lightweighted part, particularly when the internal material is removed in favour of a lightweight support structure such as a lattice. The unsupported regions deflect slightly when a polishing pressure is applied, resulting in an uneven optical surface. Research by  \textit{Atkins, C., et al. (2019b)} \cite{Carolyn2019/2} investigated using topology optimisation and various types of lattice to lightweight a mirror, using Finite Element Analysis (FEA) to aid in determining the optimal type of lattice. The paper by \textit{Westsik, M., et al. (2023)} \cite{Marcell} then took this a step further, exploring other lattices and the effect that varying the thickness had on the surface form error of a mirror, both through FEA and measurements taken from printed prototypes.

Print-through is not the only issue encountered with the polishing step of post-manufacture. One of the benefits of AM lies in the ability to print complex shapes, however, this can pose challenges when using subtractive machining on printed parts to achieve a mirror surface. The parts require some way of being held firmly in place, fixing the part in place with screws is one way of achieving this through designing designated points to attach to a jig. Screwing into the part however, can induce internal stresses that too can affect the mirror surface. \textit{Paenoi, J., et al. (2022)} \cite{NARIT} investigated this aspect of the process and in this paper two different design approaches were examined to isolate the screwing force from the optical surface; a larger `mushroom design' and a smaller design with `flexible pads'. FEA was conducted on the designs and once the designs were printed, form error and surface roughness were measured.

These are a subset of the challenges that are present in the AM process for mirrors and all are ongoing research areas, the review paper \textit{Zhang, L. et al (2021)}\cite{ZhangKai} covers some of these challenges, and others, in more detail. 
The goal of this paper will be to investigate how knowledge of the AM process for mirrors, and the issues and requirements that come from it, can be input early in the design process to to counter the negative effects. Design challenges such as reducing print-through and isolating mounting forces will therefore be explored and will build on the previous work. AM process-dependent variables such as porosity, machine settings or material choice shall remain out of scope for this project. The form this project will take is the re-design of a mirror for a deployable nanosat project called Active Deployable Optics Telescope (ADOT), details of which can be found below in Section \ref{sec:Defining the Mirror Design}. 
The aim of this project is to design a lightweight mirror that still interfaces with the original design and can be machined after printing, this can be summarised in three objectives:
\begin{itemize}
    \item Objective \#1 - Internal lattice
    \item Objective \#2 - Mounting points for machining
    \item Objective \#3 - Part consolidation of interface features
    \end{itemize}
The first objective was to lightweight the mirror by inserting a lattice into the main body. Different lattice configurations were investigated to determine the optimal support for the optical surface. This can be found in Section \ref{sec:Internal Lattice Research}. 
The second objective describes custom mounting points that were added to the mirror to allow the part to be machined, designed to limit internal stresses in the mirror surface. The third objective was to consolidate the 19 parts that make up the interface features down to one, utilising topology optimisation. Objectives 2 and 3 are discussed in Section \ref{sec:Designing Mounting Points for SPDT} and Section \ref{sec:Topology Optimisation of the Nanosat Interface Features} respectively. 
The design for AM and the manufacture and machining of the prototype are presented in Sections \ref{sec:5} and \ref{sec:6}. Surface roughness measurements are described in Section \ref{sec:7} and the paper concludes with a summary and future work in Section \ref{sec:8}

\section{Defining the Mirror Design}
\label{sec:Defining the Mirror Design}

\begin{figure}
    \centering
    \includegraphics[width = 0.5\textwidth]{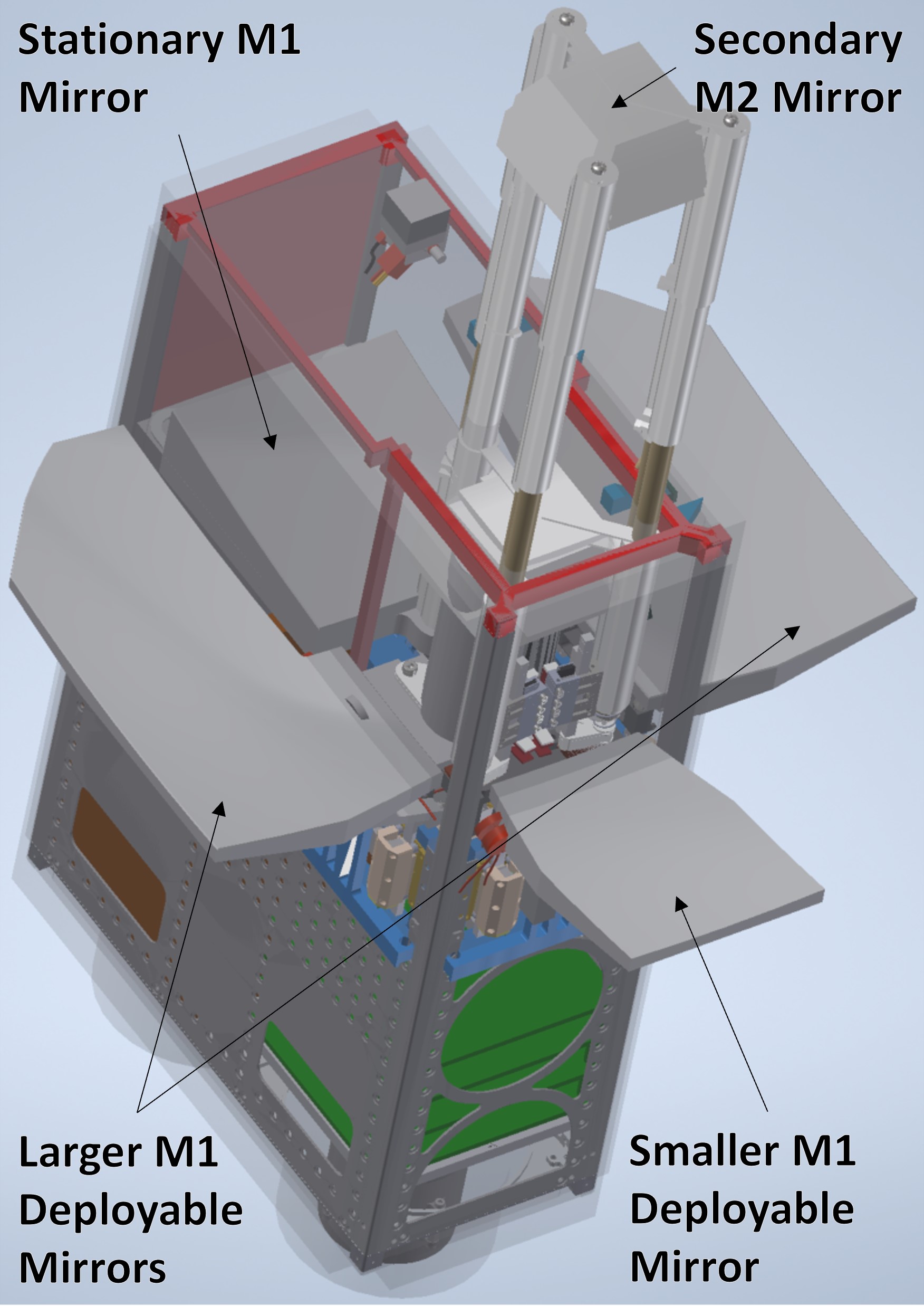}
    \caption{CAD view of ADOT (deployed) with mirrors highlighted}
    \label{fig:2}
\end{figure}

The nanosat project ADOT was selected as the base design for this study. 
ADOT is the latest iteration of a 6U nanosat, designed to specifically accommodate deployable mirrors. The paper by \textit{ Schwartz, N., et al. (2022)} \cite{Noah_ADOT} describes more detail about an earlier design iteration and highlights, ``The objective of this technology is to provide very high-angular resolution imaging in the visible at low-cost"\cite{Noah_ADOT}. The relatively small dimensions of the mirror allowed for more prototyping and suitable complexity and a high part count also benefited from simplification and consolidation. Alongside low cost being an objective of deployable and a benefit of AM mirrors, these factors were the motivation to explore redesigning these mirrors for AM. Figure~\ref{fig:2}  shows ADOT with the mirrors deployed and highlighted, the larger M1 deployable mirror was the focus of a previous AM redesign study by \textit{Westsik, M., et al. (2023)} \cite{Marcell}, this study will instead focus on the smaller M1 deployable mirror.

\begin{figure}
    \centering
    \includegraphics[width=1\linewidth]{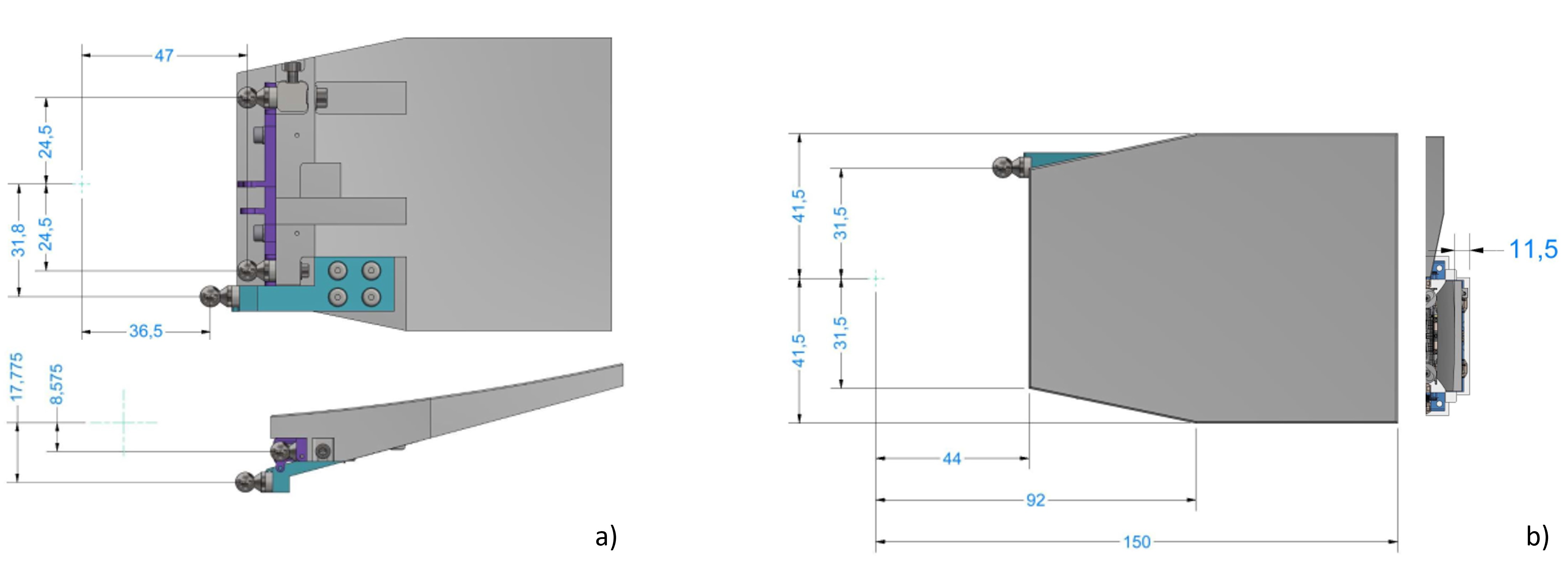}
    \caption{Mirror dimension requirements (mm): \textbf{a)} Mounting Features, \textbf{b)} Mirror external dimensions,}
    \label{fig:3}
\end{figure}

The mirror is connected to the nanosat by three tooling balls that locate on the main assembly, with seven springs that keep it in place. To ensure the mirror interfaces with the nanosat after the redesign, the important dimensions of these features had to be the same as the original. Some of the external dimensions also had to be kept to ensure the mirror would fit in the same volume within the nanosat.  These features and dimensions can be found in Table 1 and Figure \ref{fig:3}

\begin{table}
    \centering
    \begin{tabular}{|c|>{\centering\arraybackslash}p{0.6\linewidth}|} \hline 
         Mounting Features& The mirror will include three interfaces for 6 mm tooling balls\\
         &The spring attachments on the turquoise and purple components (Figure 3a) will be incorporated.\\ \hline 
         Mirror external dimensions& Optical prescription: spherical, concave\\ 
        &Spherical Radius Of Curvature (ROC): 682.28 mm\\
        &Other external dimensions are seen in Figure 3b\\ \hline
    \end{tabular}
\caption{Table showing dimensional requirements of smaller M1 mirror}
    \label{Table 1}
\end{table}

Producing a mirror in metal through the AM process comes with several requirements. The L-PBF process is used to print the mirrors in this study due to its high availability and track record with printing mirrors in previous studies, as well as its ability to print intricate geometry, high resolution and ease of powder removal. This process does come with constraints and considerations though, for example, the majority of L-PBF printers have a minimum feature size of roughly \SI{1}{\mm} that can be printed accurately. Powder removal needs to be considered as well, once a part is printed any powder that has not been melted needs a way of escaping the part, especially if the region is going to be enclosed. Finally, parts will require some supporting material to be printed, like scaffolding, to support any geometry that has large overhanging areas. This is deemed as anything that has an angle of more than 45° from the build direction. This leads to considerations such as whether parts can be designed and orientated to mitigate overhangs and can necessary supports be easily removed after printing. More information regarding design for AM can be found in \textit{``A2IM: Cookbook An introduction to additive manufacture for astronomy "}\cite{Cookbook}. 

Just as there are requirements for printing, there are also requirements for the post-manufacturing process. The main one is to determine a way of holding the part in place for the polishing process. In this study, the process will be Single Point Diamond Turning (SPDT). This is a process in which a diamond-tipped tooling bit removes nanometers of material at a time as a part rotates on a spindle, more information about the process can be found in \textit{Hatefi, S., et al (2020)}\cite{SPDT}.

The final set of requirements for this study are the self-imposed objectives. Firstly the part consolidation from 19 parts down to just one (excluding the tooling balls). Secondly, the overall Mass Reduction (MR) was targeted. The aim is to produce two different designs, one more conservative, with 50\% of the original mass of the mirror and one, more ambitious, with 70\% of the original mass.

\section{Internal Lattice Research}
\label{sec:Internal Lattice Research}

Selecting a lattice that would infill the internal region was the first stage. There are many different types of lattice to choose from, as well as other variables associated with them. So a series of experiments were conducted to ascertain which lattices and other variables would yield the best results, meeting the MR objectives and minimising form error when pressure is applied. The variables investigated were:
\begin{itemize}
    \item Lattice type
    \item Lattice size
    \item Blend radius
    \item Lattice thickness
    \item Cell map
\end{itemize}
Each was investigated according to the anticipated impact and time available. To start, results were reviewed from the paper by \textit{Westsik, M., et al. (2023)} \cite{Marcell}, as the application was very similar and the same AM software was used. The software used was called nTopology and the down-selection process was taken from the library of lattices available on the platform.

\subsection{Down-selecting the Lattice Type}
\begin{figure}
    \centering
    \includegraphics[width=0.5\linewidth]{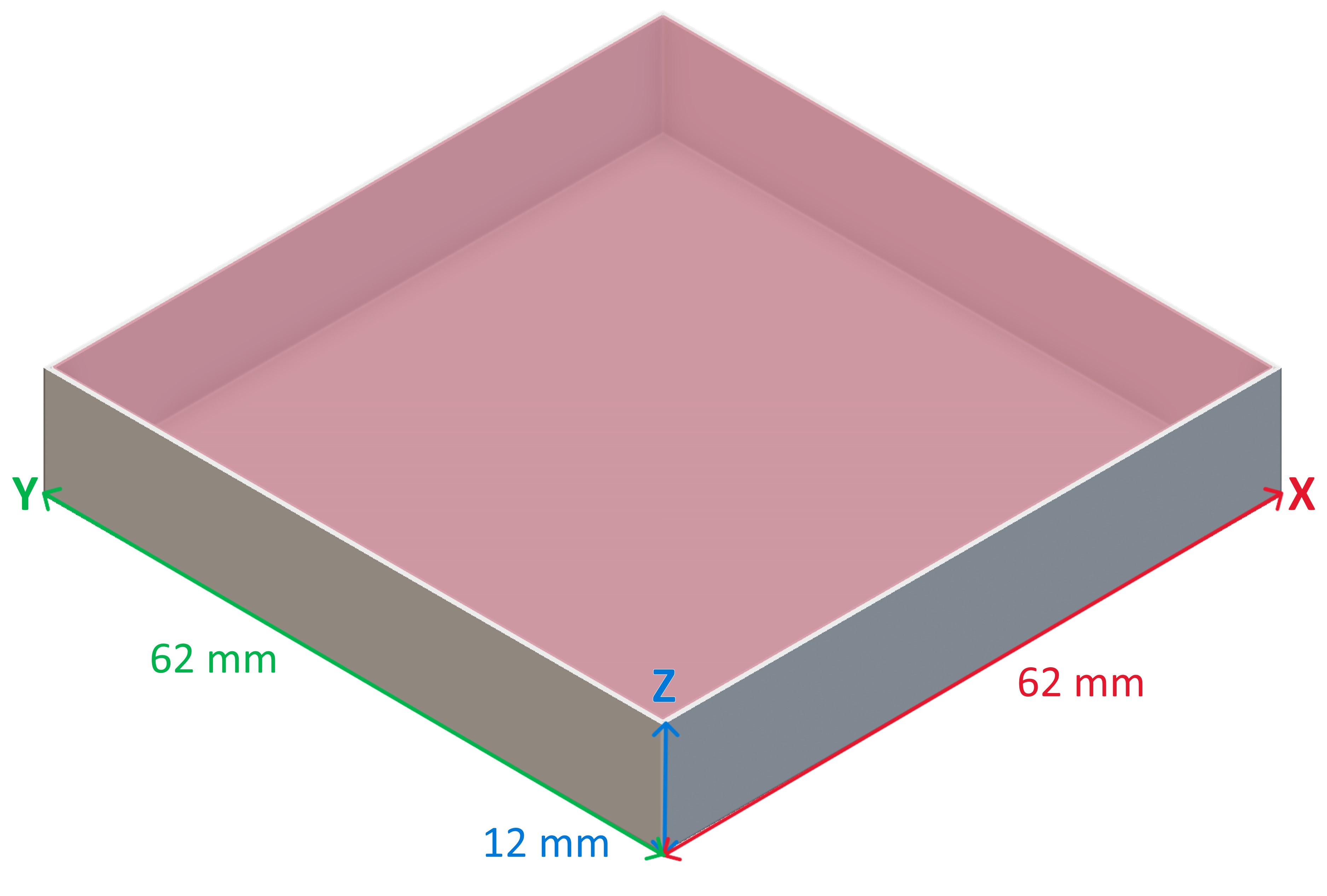}
    \caption{Sample box shape used for lattice experiments}
    \label{fig:4}
\end{figure}

To conduct these initial experiments a sample box shape was used as a control volume to give a constant simple shape between each variable and allow for faster processing speed. The box can be seen in Figure \ref{fig:4}, with the top surface removed and the internal region (red) depicting the area for a lattice to fill. The dimensions can be seen in the figure, the lattice volume is 60 x 60 x \SI{10}{\mm}, surrounded on six sides by a shell, with a thickness of \SI{1}{\mm}.

\begin{figure}
    \centering
    \includegraphics[width=1\linewidth]{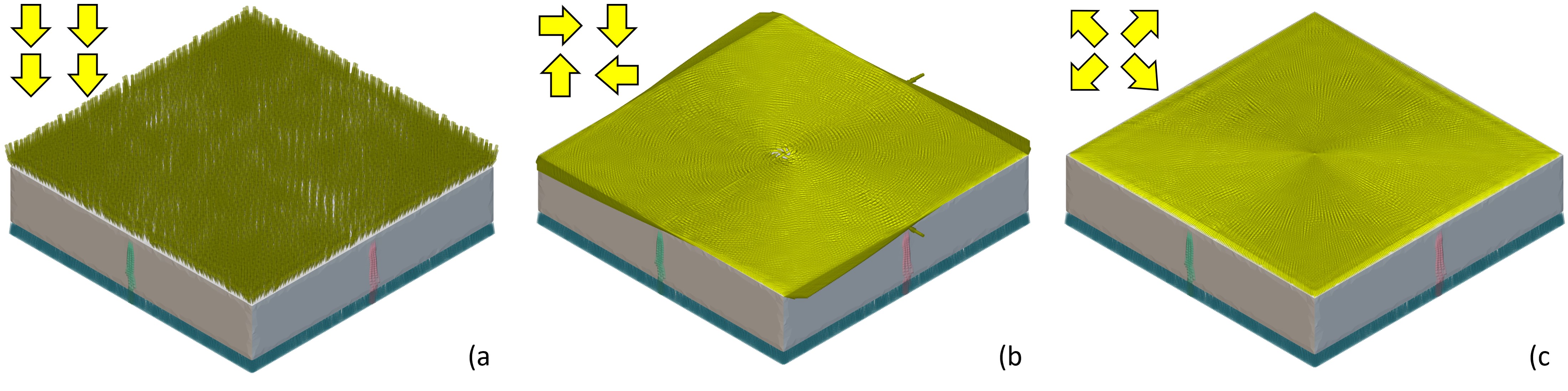}
        \caption{Load Cases for Lattice Experiments: \textbf{a)} Direct Pressure Load, \textbf{b)} Circumferential Force,\textbf{ c)} Radial Force,}
    \label{fig:5}
\end{figure}

Three different load cases were created to roughly replicate the type of loads that the mirror could experience in the machining processes, depicted in Figure  \ref{fig:5}, these were a direct load, a circumferential load and a radial load. In each case, the base was fixed in the Z direction and the centre of each side (X \& Y) was fixed in the direction parallel to it. The magnitude of the pressure was chosen at  \SI{3500}{\Pa}, an arbitrary approximation to the polishing pressure, so comparisons between results could be obtained. The value has also been used in previous studies\cite{Marcell,CAtkins2017,C.Atkins-2019-01,Carolyn2019/2,Roulet2018,NARIT}.

\begin{figure}
    \centering
    \includegraphics[width=0.75\linewidth]{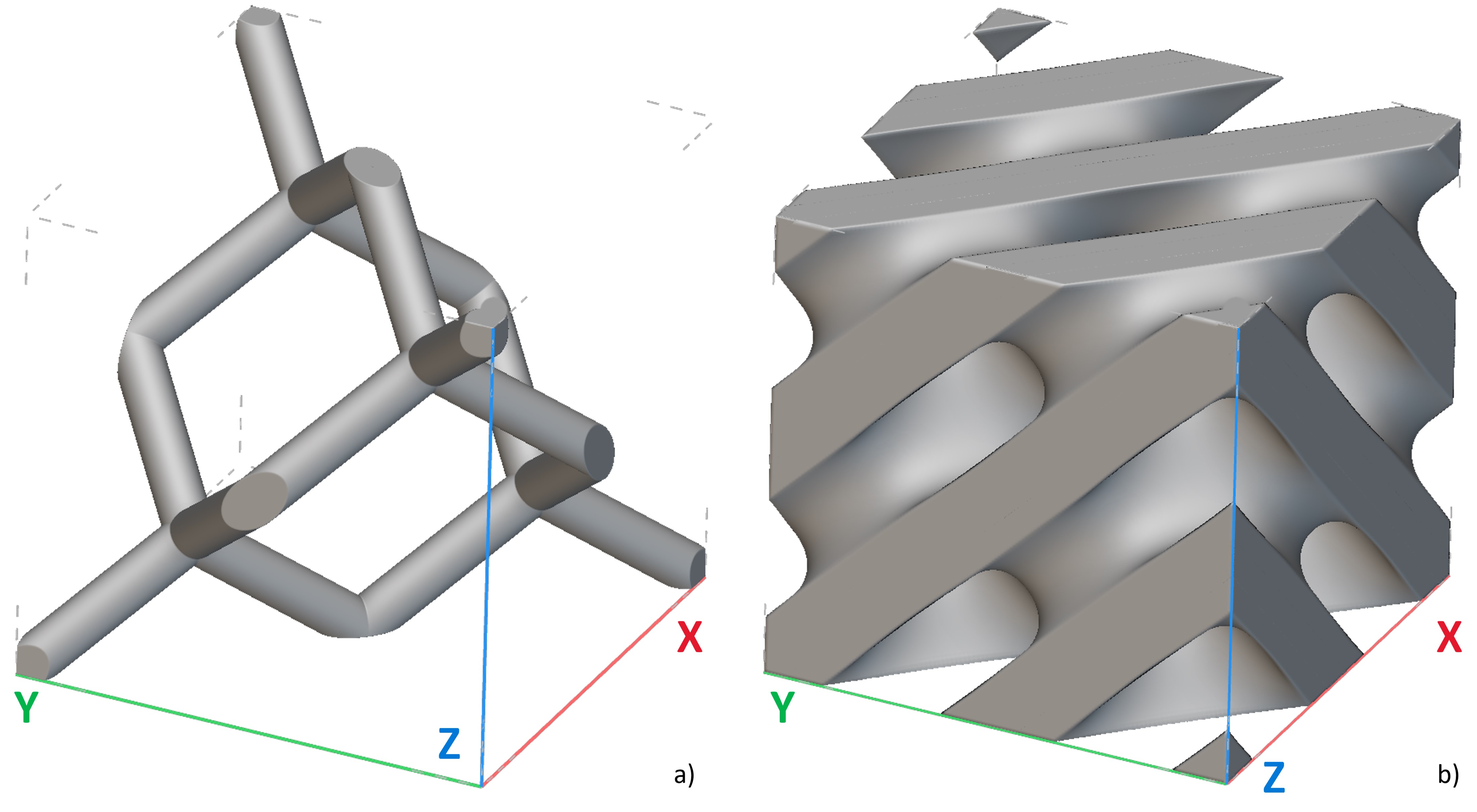}
    \caption{Unit Cells \textbf{a)} Graph - Diamond, \textbf{b)} TPMS - Diamond}
    \label{fig:6}
\end{figure}

\begin{figure}
    \centering
    \includegraphics[width=0.75\linewidth]{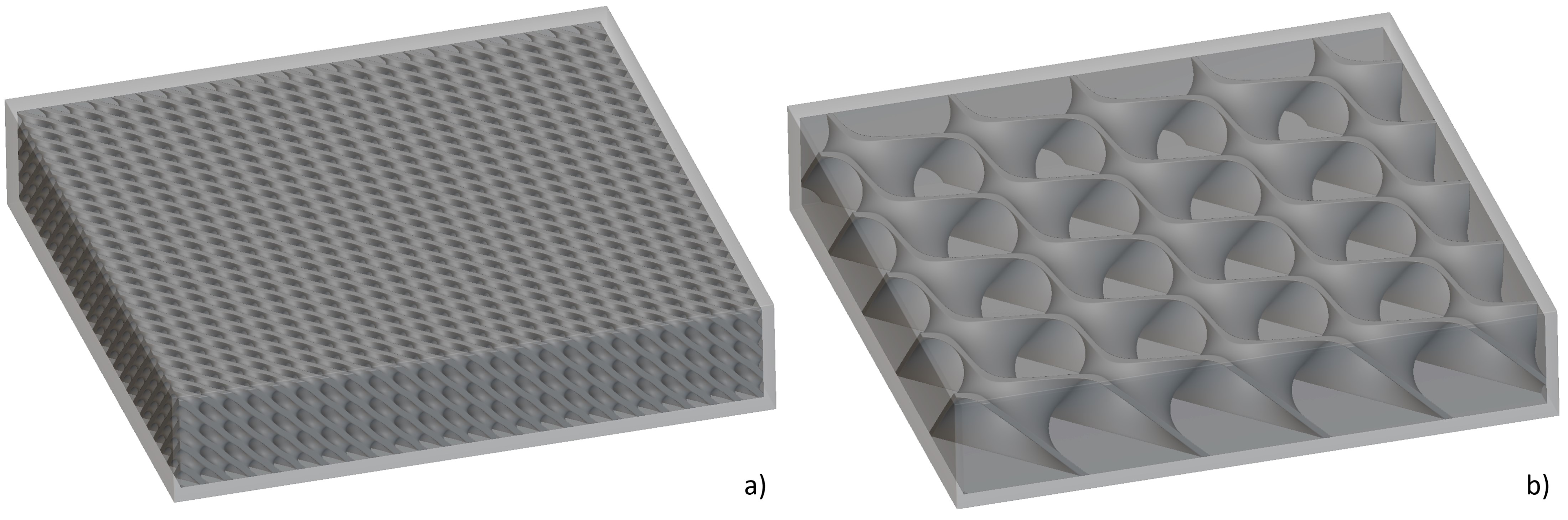}
    \caption{Varied Cell Sizes \textbf{a)} 5 x 5 x \SI{5}{\mm} Diamond TPMS, \textbf{b)} 25 x 25 x \SI{25}{\mm} Diamond TPMS}
    \label{fig:7}
\end{figure}

The first two variables changed were the type of lattice inside the shell and the size of the lattice unit cell. Two types of unit cells will be considered in this study, graph unit cells and Triply Periodic Minimal Surface (TPMS) unit cells, an example of both can be seen in Figure \ref{fig:6}. Graph-based lattices are ``composed of a group of crossbars (s) interconnected with each other in points called nodes (n)"\cite{Unitcells} and TPMS-based lattices are ``generated by algorithms and can be represented by mathematical equations"\cite{Unitcells}. More information on both types can be found in the article \textit{Riva, L., et al. (2021)}\cite{Unitcells}.  The results of  \textit{Westsik, M., et al. (2023)} \cite{Marcell} found that ``TPMS lattices perform better than, or identically to, the graph lattices for the same WR values"\cite{Marcell}, as such the majority of unit cells tested in this study were TPMS, with some graph unit cells included that also performed well. However, in these experiments, the size of the unit cell would be the variable. The distances of X, Y and Z seen in Figure \ref{fig:6} can all be varied to stretch the size of the unit cell and therefore also the number of cells in a volume, in this figure the dimensions are equal, resulting in a cube shape. 
Figure \ref{fig:7} shows this in a lattice, the first being more dense, where each cell is a cube in dimensions 5 x 5 x \SI{5}{\mm}, and the second being more lightweight, with each cell being \SI{25}{\mm} in each direction.
The size of each type of unit cell cube was varied between the dimensions of \SI{2.5}{\mm} to \SI{25}{\mm} to achieve both useful and realistic MR results.

The MR would then be plotted against the performance of each type. The performance was quantified by measuring the displacement of the top \SI{1}{\mm} of the surface after the pressure was applied. Only the direct pressure was used during these initial tests due to the large volume of experiments and the limitations of the software. Displacement was measured in two ways, Peak to Valley (PV) and Root Mean Square (RMS). PV is calculated by subtracting the displacement minimum from the maximum, and RMS is measured by squaring all displacements, finding the mean and then square-rooting the mean. 
\begin{figure}
    \centering
    \includegraphics[width=0.75\linewidth]{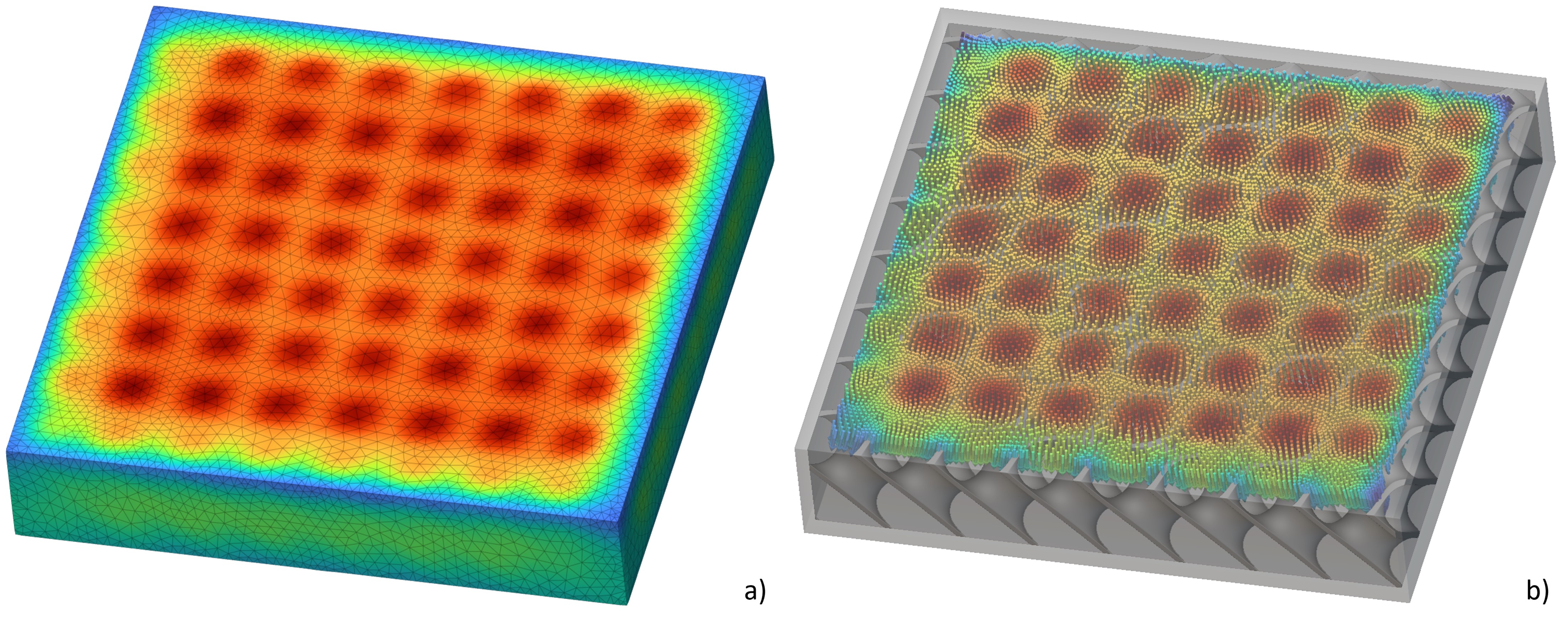}
    \caption{Sample results from varying cell size:\textbf{ a)} full stress analysis showing displacement, \textbf{b)} top region isolated to obtain PV and RMS from data points }
    \label{fig:8}
\end{figure}
Figure \ref{fig:8} shows the displacement on a 15 x 15 x \SI{15}{\mm} lattice and the set of points that were selected to analyse for performance. The area of points that were used were any that were within 1mm from the top surface and at least 3mm from the edge. This represents the deformation on the top surface of the mirror during machining and reduces the edge effect in providing extra support. The walls on the side were however necessary to support the lattice and reduce erroneous results, where a lattice wouldn't support the top surface and lead to an exaggerated displacement result.
\begin{figure}
    \centering
    \includegraphics[width=1\linewidth]{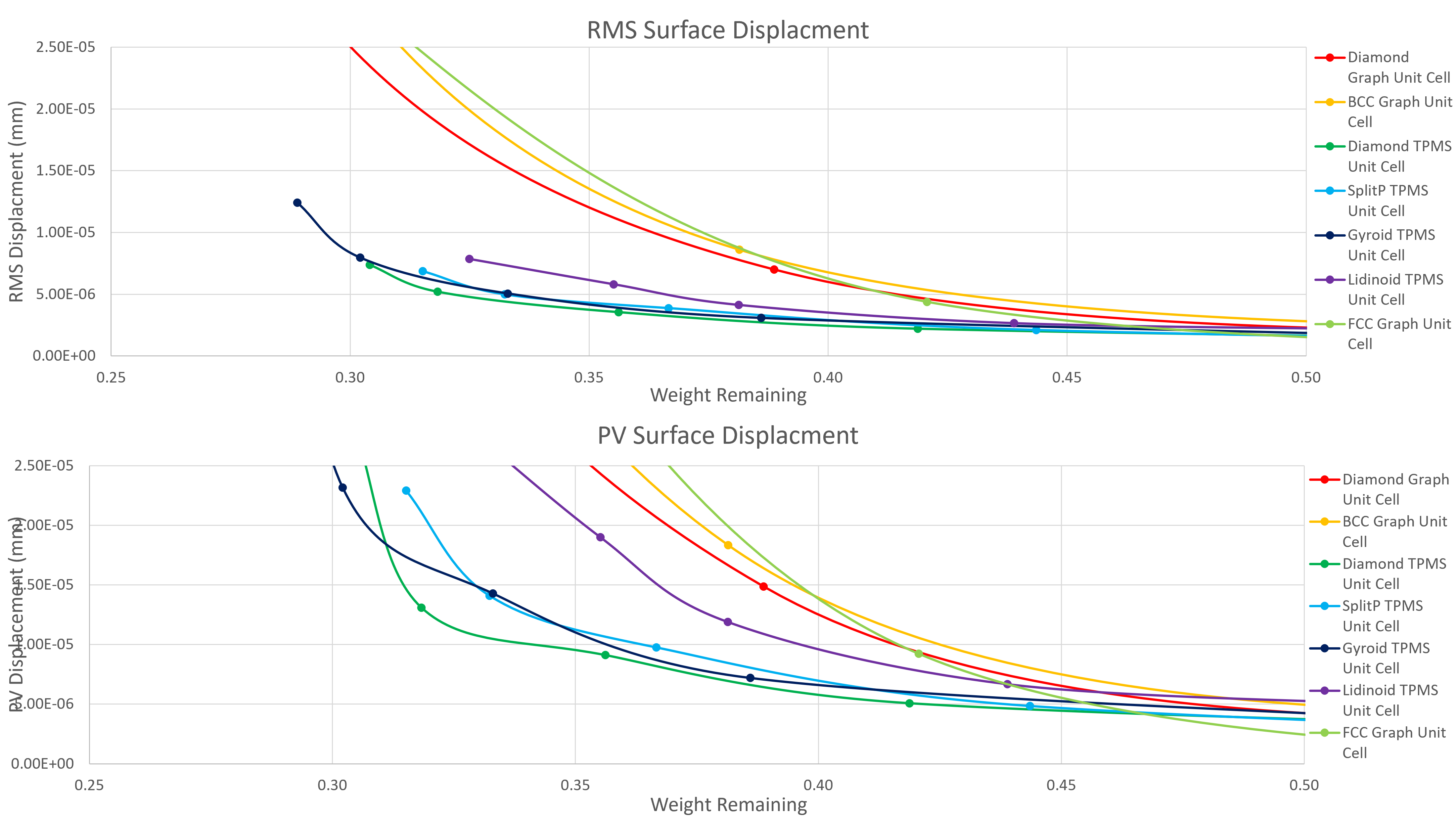}
    \caption{Results from varying cell size and lattice type, in these plots weight remaining is defined equal to 1 - the mass reduction.}
    \label{fig:9}
\end{figure}

From the results in Figure \ref{fig:9}, the Diamond TPMS Lattice performed the best, yielding the lowest weight remaining while maintaining a relatively low PV and RMS displacement. The lower weight remaining is given by larger dimensional unit cells, the larger the size of unit cells the less number of cells in the volume. The two points that yielded the best results were the 15 and 20 mm cubes, especially for the Diamond TPMS. These two points give a rough MR of 64\% and 68\% respectively.

\subsection{Experimenting with Stretching the Lattice}
\label{sec:Experimenting with Stretching the Lattice}
\begin{figure}
    \centering
    \includegraphics[width=0.75\linewidth]{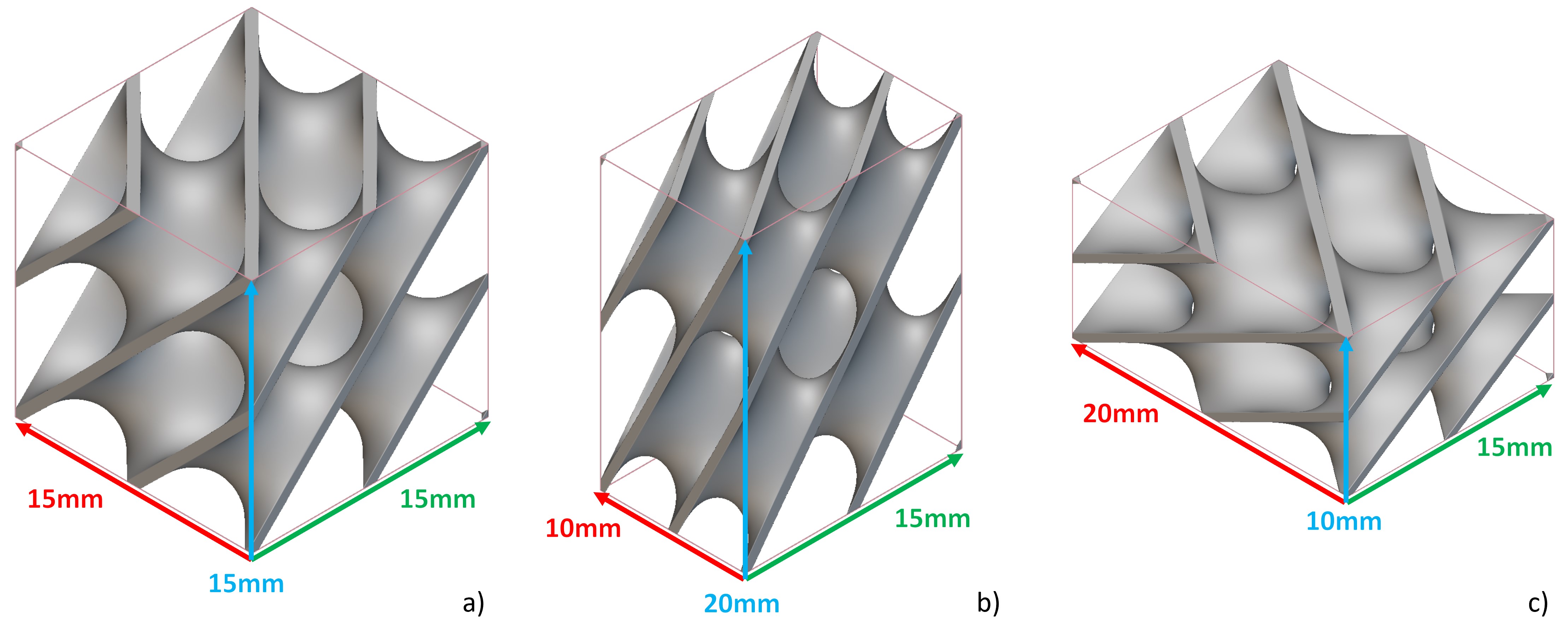}
    \caption{Diamond TPMS unit cells stretched in X $\times$ Y $\times$ Z: \textbf{a)}  15 $\times$ 15 $\times$ 15, \textbf{b)}  10 $\times$ 15 $\times$ 20, \textbf{c)}  20 $\times$ 15 $\times$ 10}
    \label{fig:10.5}
\end{figure}

\begin{figure}
    \centering
    \includegraphics[width=0.75\linewidth]{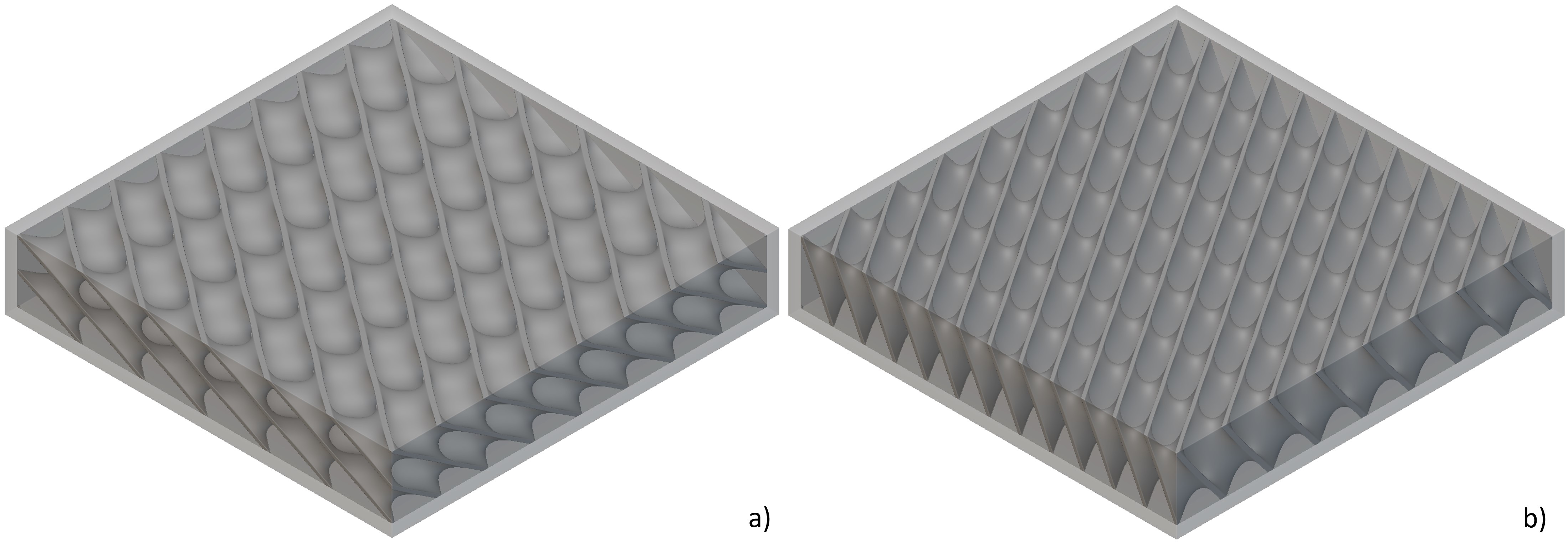}
    \caption{Diamond TPMS Lattice with stretched unit cells: \textbf{a)} 20 $\times$ 15 $\times$ 10, \textbf{b)} 10 $\times$ 15 $\times$ 20 }
    \label{fig:10}
\end{figure}

This first experiment maintained the cube shape of the unit cell, varying the dimensions X, Y \& Z evenly. However, each edge length can be changed independently, stretching the unit cell into a cuboid. The next experiment investigated the effect stretching the unit cell has on performance, the Diamond TPMS was selected for this experiment as it performed the best in the previous test, similarly, the MR of 64\% was chosen to keep constant. The original cube unit cell for this value was  15 $\times$ 15 $\times$ \SI{15}{\mm}, so each side was compressed and stretched to maintain the same volume but with a different shape, giving each unit cell dimensions of 10, 15 and 20. The cuboid unit cells with different lengths can be seen in Figure \ref{fig:10.5}.
An example of two of the shapes tested is seen in Figure \ref{fig:10}, the rest of the experimental set-up was the same as the first and the original cube unit cell was included for comparison.

\begin{figure}
    \centering
    \includegraphics[width=1\linewidth]{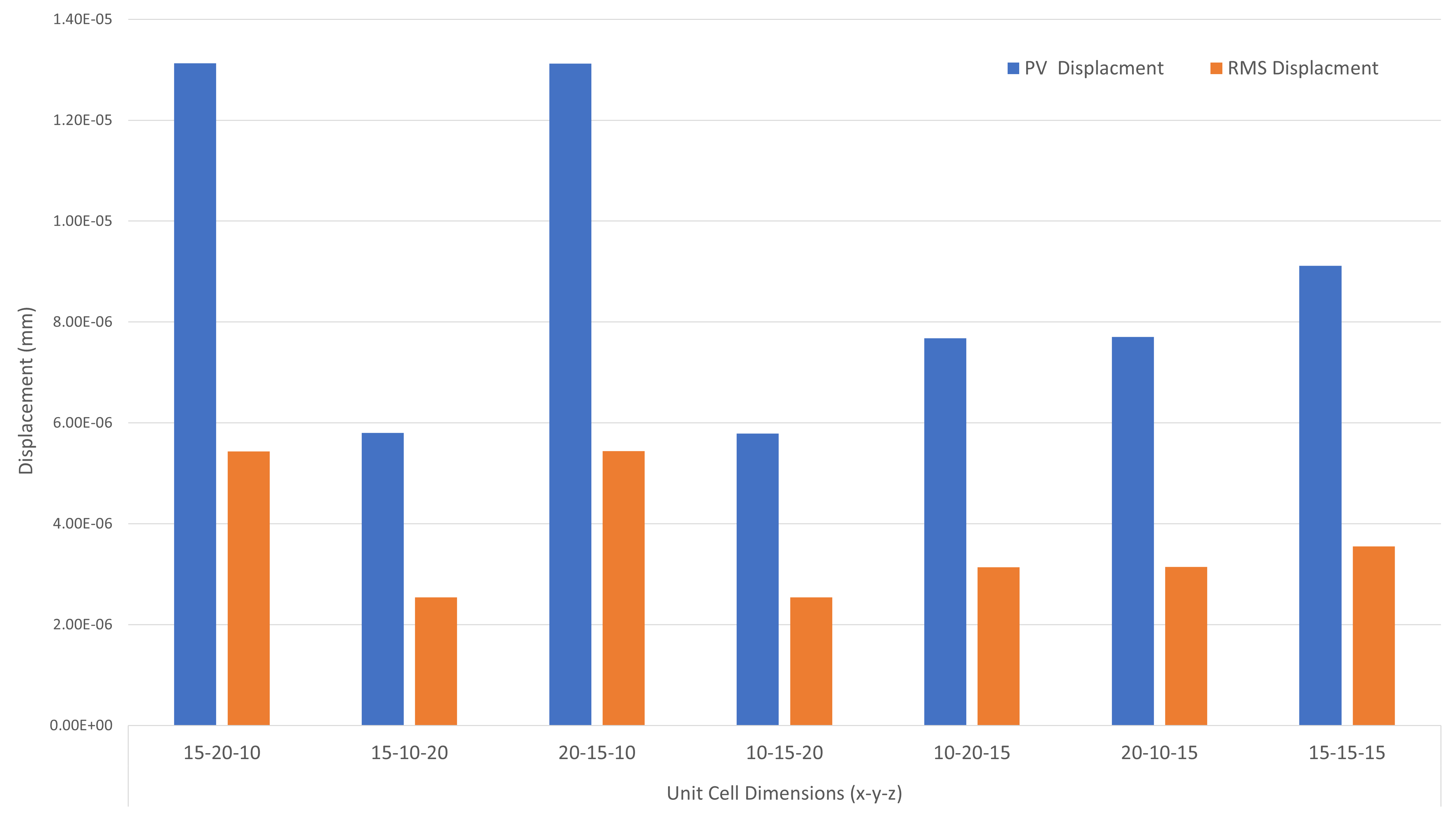}
    \caption{Displacement results of stretching Diamond TPMS unit cells under pressure}
    \label{fig:11}
\end{figure}

The result displayed in Figure  \ref{fig:11} suggests that stretching the unit cells vertically yields lower displacements. The stretched unit cell that performed best, 10-15-20, can be seen in Figure \ref{fig:10.5}b. Stretching in this way allows more cells in the X and Y directions and more contact points with the top surface. It is also worth noting that the unit cell can only be stretched so far before the geometry is no longer suitable for printing, a requirement considered throughout the process.

\subsection{Investigating the Effect of Blend Radius}
\begin{figure}
    \centering
    \includegraphics[width=1\linewidth]{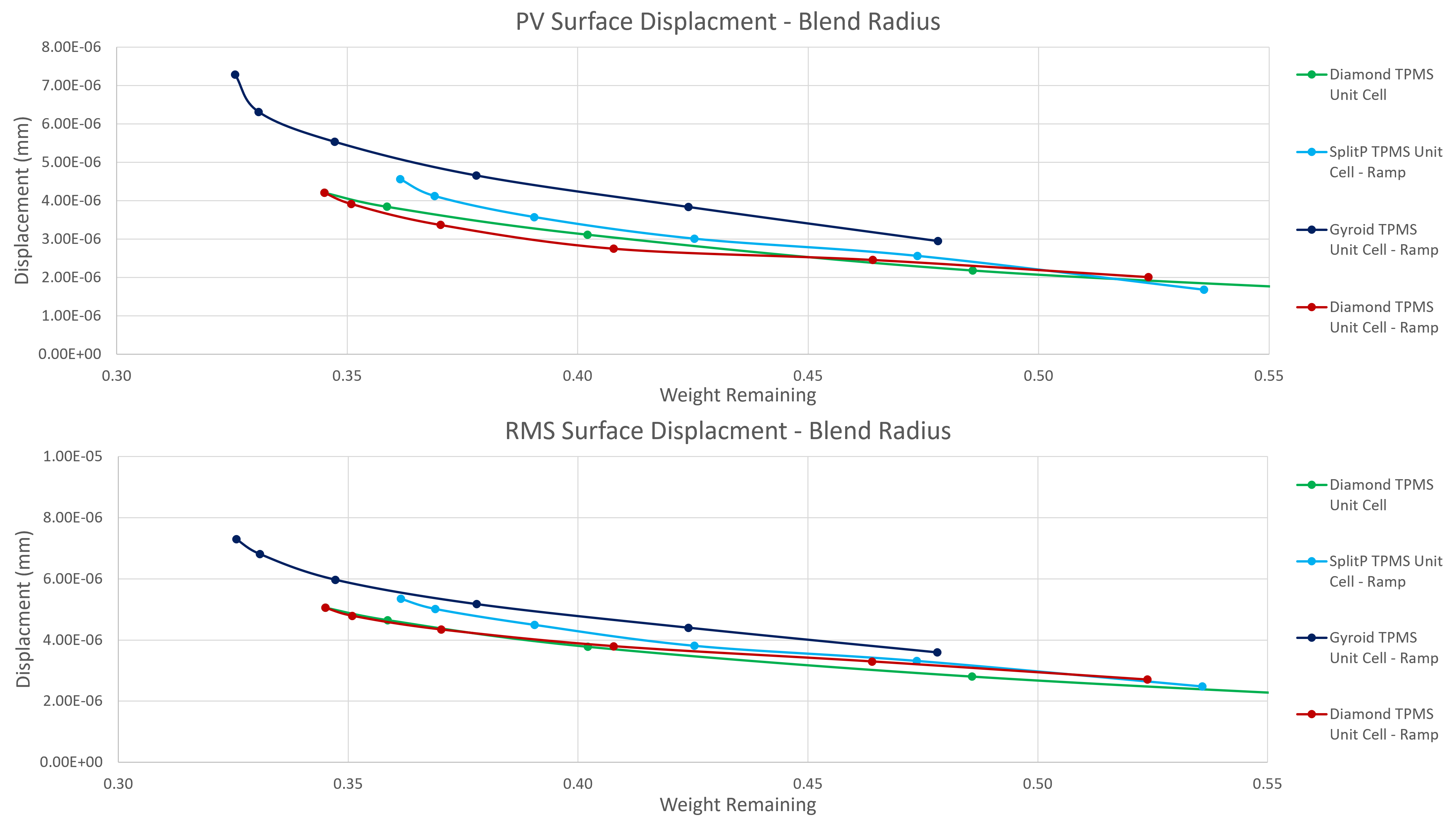}
    \caption{Results from Varying the Blend Radius}
    \label{fig:12}
\end{figure}

When joining two different bodies in the AM software, it is possible to add a radius between the contact edges, to create a rounded transition. It was considered that this may add increased support to the top surface, without adding too much extra weight. As such, an experiment on different radius sizes was conducted ranging from 0 to \SI{5}{\mm}, with smaller radii adding less material and giving smaller weight remaining results. The option to add a `ramp' to this radius was available as well which increases the thickness closer to the top surface. The results are shown in Figure  \ref{fig:12}.
From the results, we can see the Diamond TPMS once again performing the best when compared to the other TPMS lattices tested. The `ramp' feature can be said to have a marginal improvement in performance over the smaller radii but this deteriorates with more material added. The conclusion from this experiment was that the blend radius has a small benefit but can quickly add lots of mass with larger radii. A 1 mm blend radius was selected to reduce the impact on the mass and to smooth the interface between the lattice and the surface, which is a common location of stress during printing.

\subsection{Unit Cell Selection}
Examining the results from all the experiments, the Diamond TPMS lattice and the Diamond Graph lattice were selected for the designs. The TPMS was to be used in the 70\% MR design due to its good performance at low-weight remaining experiments and its history of printability\cite{Marcell}. However, with such a stretched shape to achieve the high MR, there were still concerns about how well it would print. It was therefore decided that the Diamond Graph unit cell would be used in the 50\% MR design, as it performed well compared to the other graph cells and was more comparable at 0.5 weight remaining to the TPMS cells, further, it had been successfully used in previous studies (print-ability).

\begin{figure}
    \centering
    \includegraphics[width=1\linewidth]{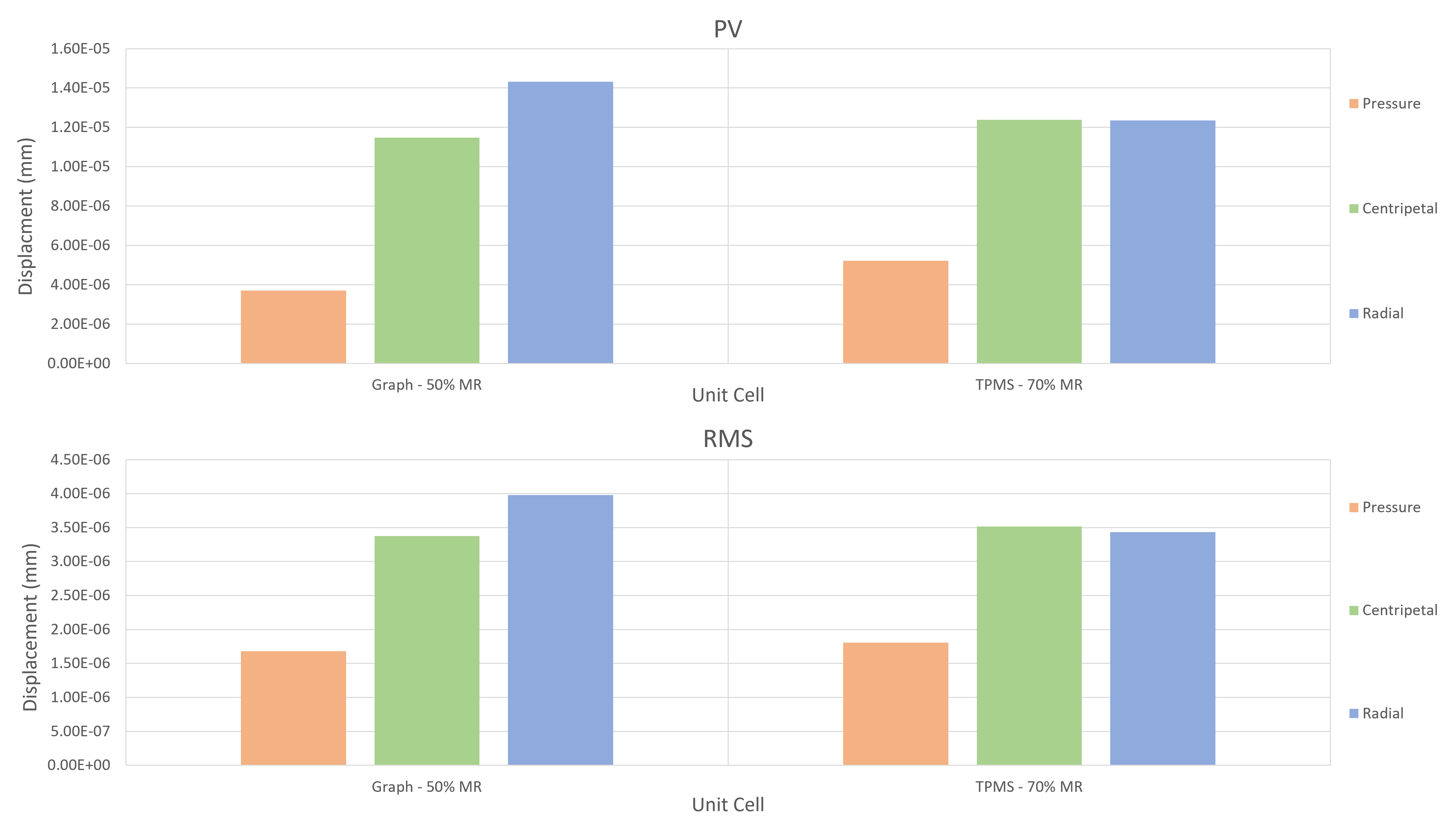}
    \caption{Results from three load cases on Diamond Graph and Diamond TPMS lattices}
    \label{fig: 13}
\end{figure}

To further validate these two types of lattice, they underwent the other two load cases seen in Figure  \ref{fig:5}, the circumferential load and a radial load. The density of the lattice was selected to replicate the same MR as the set objective. The results can be seen in Figure  \ref{fig: 13}.
Comparable results between the two types of lattice can be seen for each load case, despite the TPMS having a much larger MR of 70\%. The calculated displacement values were not considered as the load was the same as the one used in the pressure case, an arbitrary value. Further research could look into a more accurate set of boundary conditions for SPDT to be used that could then be applied to more types of unit cells for comparison.

\subsection{Cell Map Comparison}
\begin{figure}
    \centering
    \includegraphics[width=0.75\linewidth]{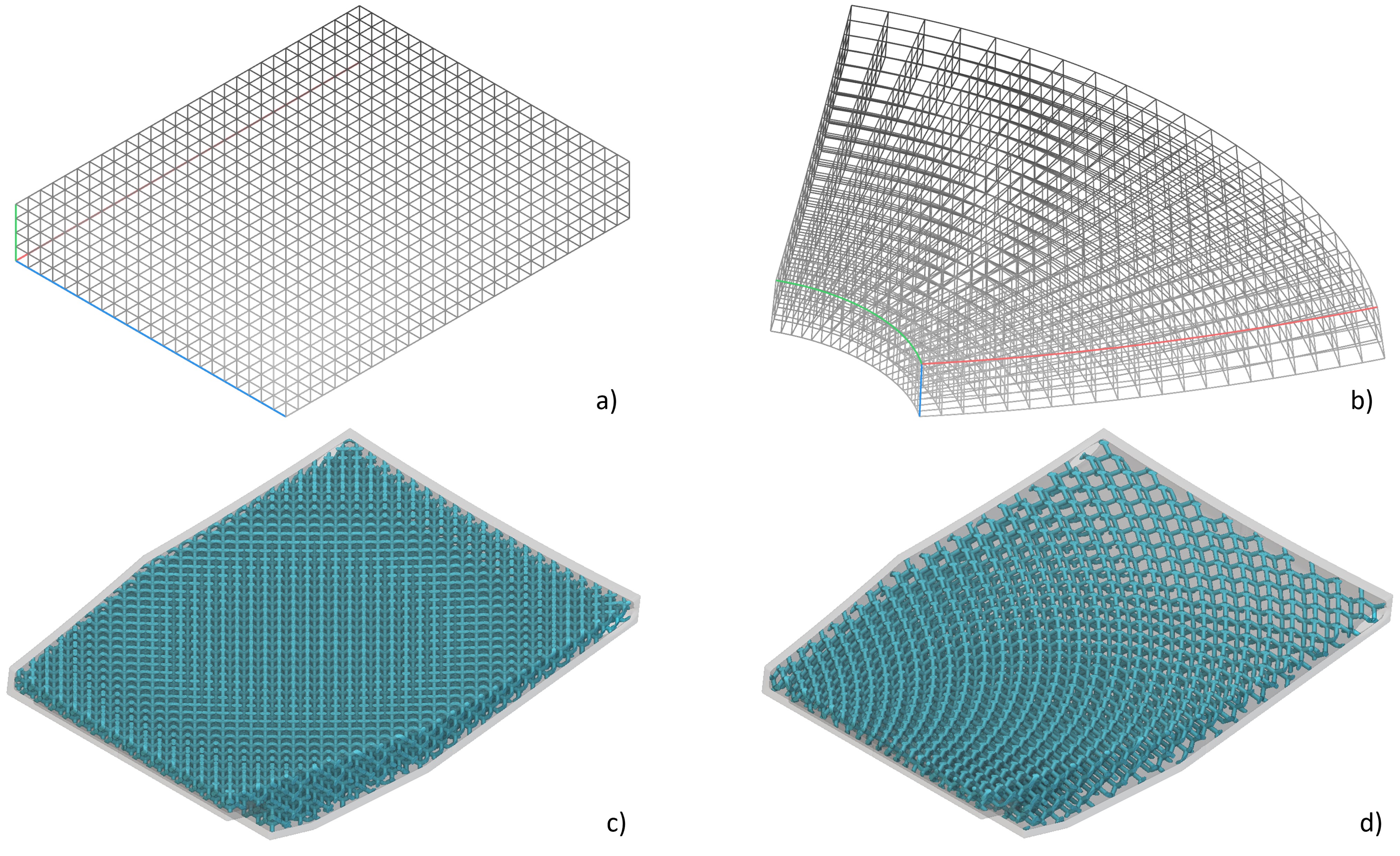}
    \caption{Different shape cell maps and lattices: \textbf{a)} Rectangular cell map, \textbf{b)} CAD surface cell map, \textbf{c)} Diamond Graph lattice mirror infill using a rectangular cell map, \textbf{d)} Diamond Graph lattice mirror infill using a cell map derived from the CAD of the mirror's surface, }
    \label{fig:14}
\end{figure}

With the lattice selected, the next step was to implement the lattice with the geometry of the mirror. The main body of the mirror was hollowed out, and grooves and holes that were subtractive features for assembly were filled in to give a more uniform shape. A lattice could then be generated for the internal volume using a cell map. A cell map defines the location of each unit cell and its shape. Until this point, a rectangular cell map that uses a Cartesian coordinate system had been used, as the test volume was rectangular. However, with the body containing a spherical mirror surface, a cell map was trialled that used the CAD from the mirror surface to construct cell map. The spherical surface gave this cell map a polar coordinate system, it was suspected that this would provide better support. Figure \ref{fig:14} shows the rectangular cell map, and resulting lattice, versus the 'mirror' cell map and lattice.
In this figure, the thickness of each strut in the lattice was consistent at \SI{1}{\mm}. The thickness of the lattice can, however, be varied throughout the volume, either by a defined function or a 2D/3D data map. The spherical nature of the CAD surface cell map shows that the density of the lattice is larger towards the front of the mirror and is not uniform like the rectangular cell map. Combining this varying density with the geometric tapering of the mirror, meant that a varying thickness lattice was used to reduce the impact of these non-uniformities. This resulted in a lattice that increased in thickness at the back of the mirror, varying in degree between the two designs. The Graph lattice increased from \SI{1}{\mm} to \SI{2}{\mm}, where as the TPMS lattice increased from \SI{1}{\mm} to \SI{1.1}{\mm}.

\begin{figure}
    \centering
    \includegraphics[width=1\linewidth]{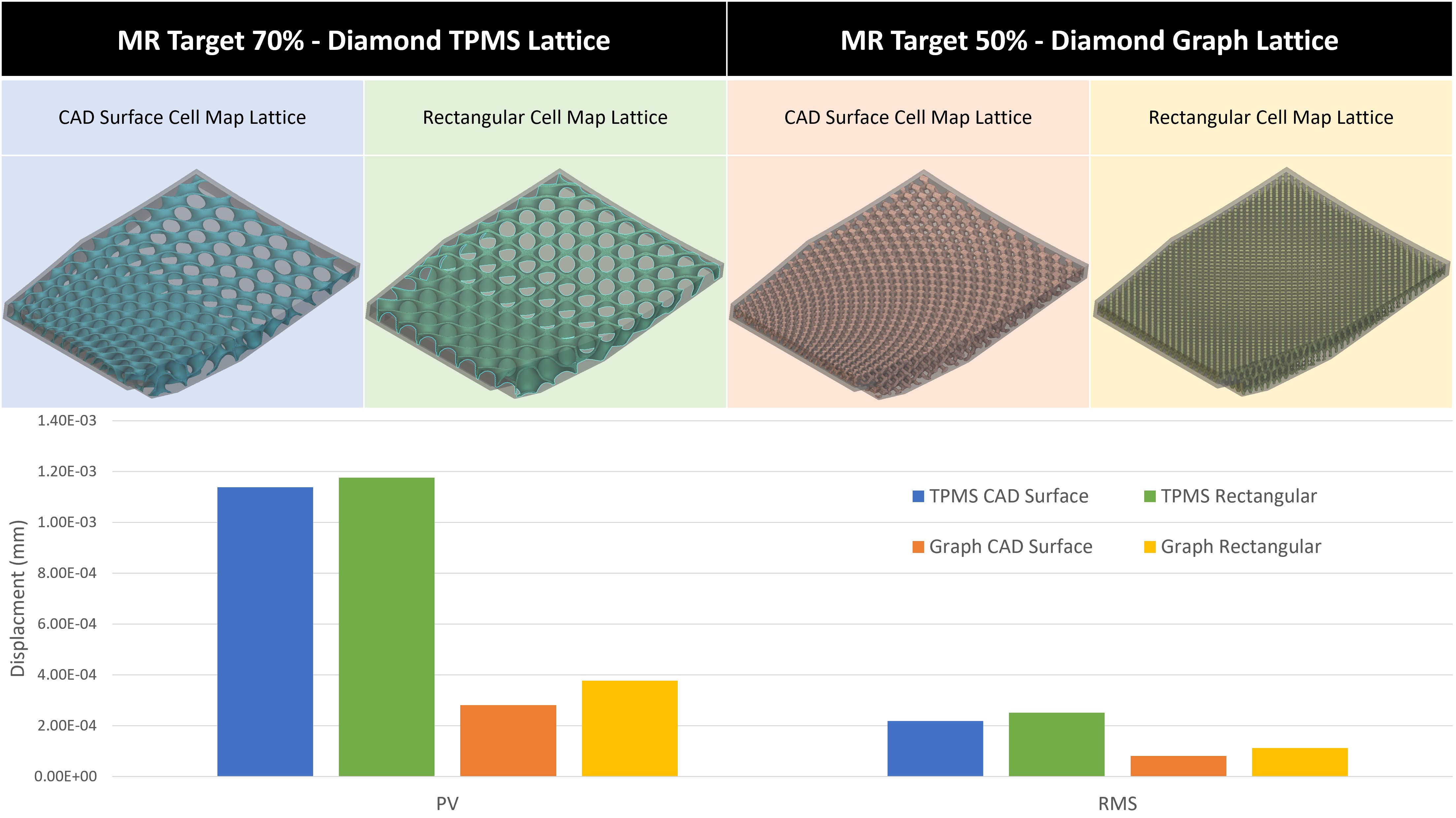}
    \caption{Four different types of lattice and cell map compared to achieve minimal surface displacement}
    \label{fig:15}
\end{figure}

\begin{figure}
    \centering
    \includegraphics[width=0.75\linewidth]{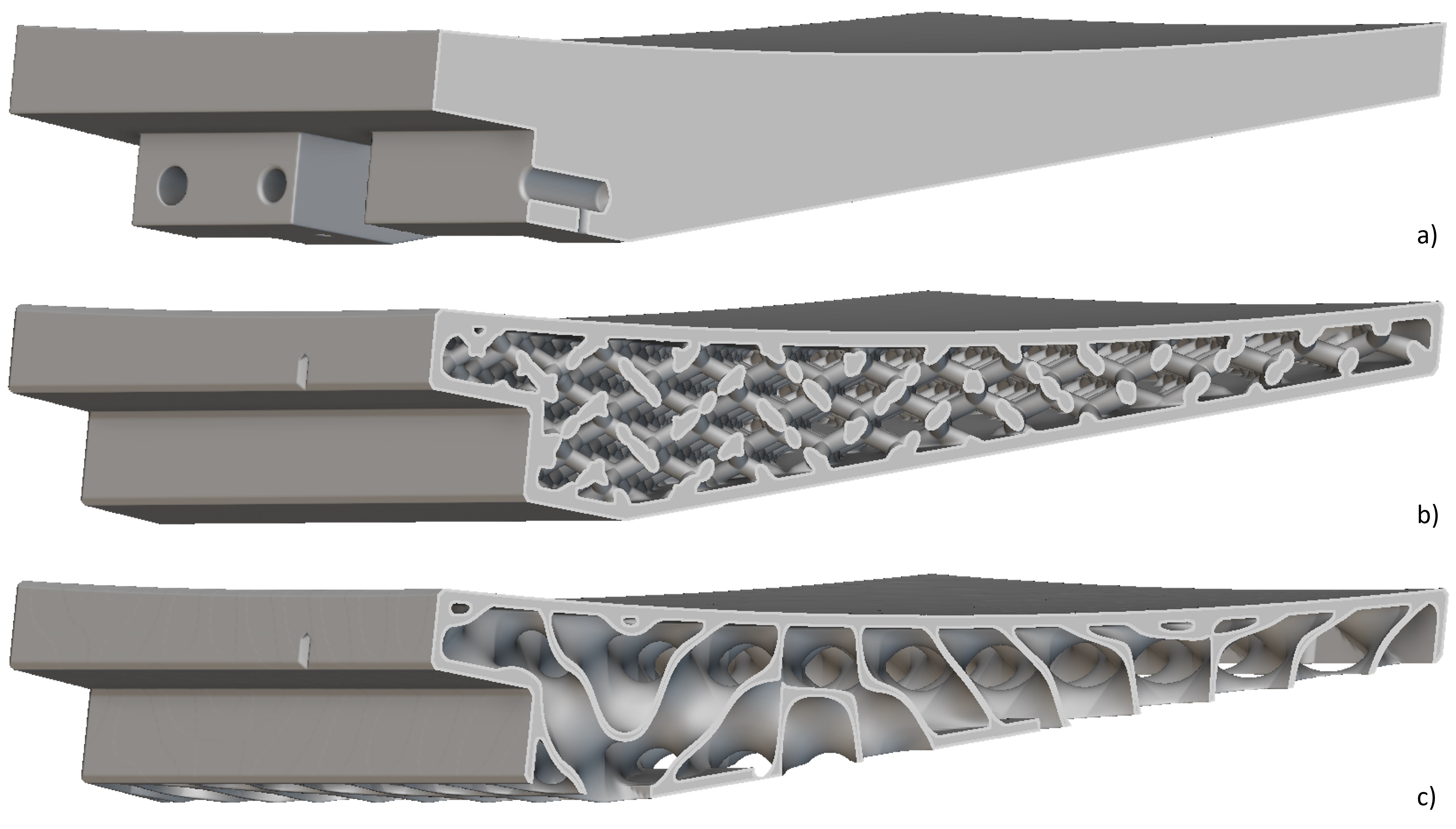}
    \caption{Internal lattice main body designs: \textbf{a)} Original main body,  \textbf{b)} Closed back main body with a Diamond Graph lattice infill on a CAD surface derived cell map, \textbf{c)} Open back main body with a Diamond TPMS lattice infill on a CAD surface derived cell map}
    \label{fig:16}
\end{figure}

To compare the two types of cell maps an experiment was set up. Four different lattice infills were created, two Graph and two TPMS, a rectangular and a CAD surface cell map version of each. Each was created by varying the density of the lattice and the thickness as stated above to achieve the MR objectives of each design. 50\% MR using the Graph lattice was achieved using this method; however, 70\% MR using the TPMS lattice was challenging due to the limitation of printability and lattice thickness. Therefore the decision was made to remove the back of this mirror, so the TPMS lattice cell map variants will feature an open-back design. Whilst this did reduce some rigidity it also removed mass and allowed for a thicker lattice to be used. 
The set-up for the design was an arbitrary pressure of  \SI{3500}{\Pa} applied to the mirror surface and the shape fixed at three mounting points (see Section \ref{sec:Designing Mounting Points for SPDT}). Each lattice infill design and the results can be seen in Figure \ref{fig:15}.
These results suggest a marginal improvement in performance in the CAD surface lattice cell map in both cases, therefore this one was selected going forward. 
The number of unit cells and therefore the density of the lattice could then be finalised, aiming to keep the geometry of the lattice printable but also make use of the findings from Section \ref{sec:Experimenting with Stretching the Lattice}, ultimately aiming for just over the MR objectives for each design. The thickness of the lattice could then be varied as described before, increasing the thickness at the thinner part of the mirror to achieve the MR objectives.
The original ADOT body design, along with the new body designs, can be seen in Figure \ref{fig:16}.

\section{Designing Mounting Points and Topology Optimisation}
\label{sec:4.}

\subsection{Designing Mounting Points for SPDT}
\label{sec:Designing Mounting Points for SPDT}

Whilst some of the geometry that is not necessary for AM will be removed, such as channels for screws,  some features will need to be added, namely, some mounting points for SPDT. The principle design for these mounting points drew inspiration from the `Flexible Pads' design that was studied in the paper \textit{Paenoi, J., et al. (2022)} \cite{NARIT}, this paper had a very similar objective in attempting to isolate forces from screwing into the part from the mirror surface. While the pads demonstrated an improvement in form error, the software available limited this study from going further, as such, this resulted in a conventional geometry which does not fully make use of the intricacies available through AM. To this end, a lattice-inspired design was created using the lightweight and flexible lattice attributes.

\subsubsection{Initial Lattice Down-Selection}
\begin{figure}
    \centering
    \includegraphics[width=1\linewidth]{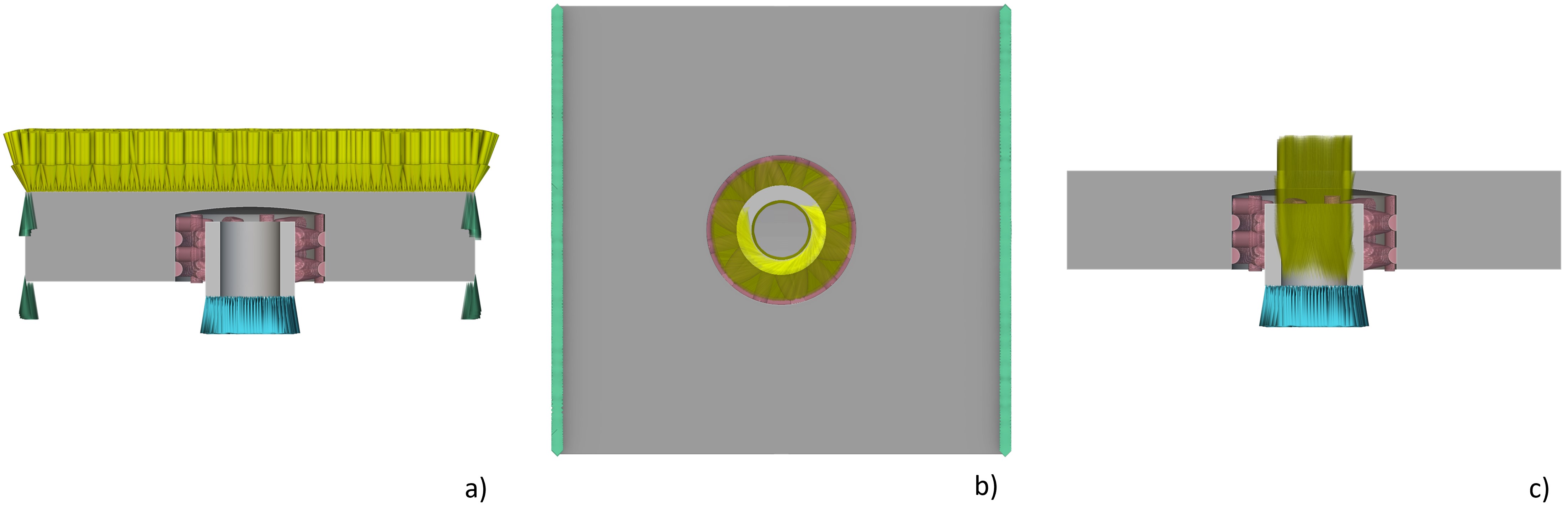}
    \caption{Load cases for lattice selection: \textbf{a)} Pressure case, \textbf{b)} Screw case, \textbf{c)} Pull case}
    \label{fig:17}
\end{figure}

Following the procedure outlined in Section \ref{sec:Internal Lattice Research}, three load cases were considered and a test volume was used, these can be seen in Figure \ref{fig:17}.
The overall shape consisted of a box, 15 x 15 x \SI{6}{\mm}, with a domed cylindrical cavity that connected to a hollowed cylinder by a lattice (red). The hollow cylinder protruded slightly and acted as the main mounting foot with the hole intended for the screw fixture. The outer box therefore represented the mirror body and the top of the box the mirror surface. In the figure, each load (yellow) represented either the SPDT pressure or the forces of screwing into the part,  the models were fixed on the base pad (blue) in the first and last case and on the the side of the box (green) in the screw case.

\begin{figure}
    \centering
    \includegraphics[width=0.5\linewidth]{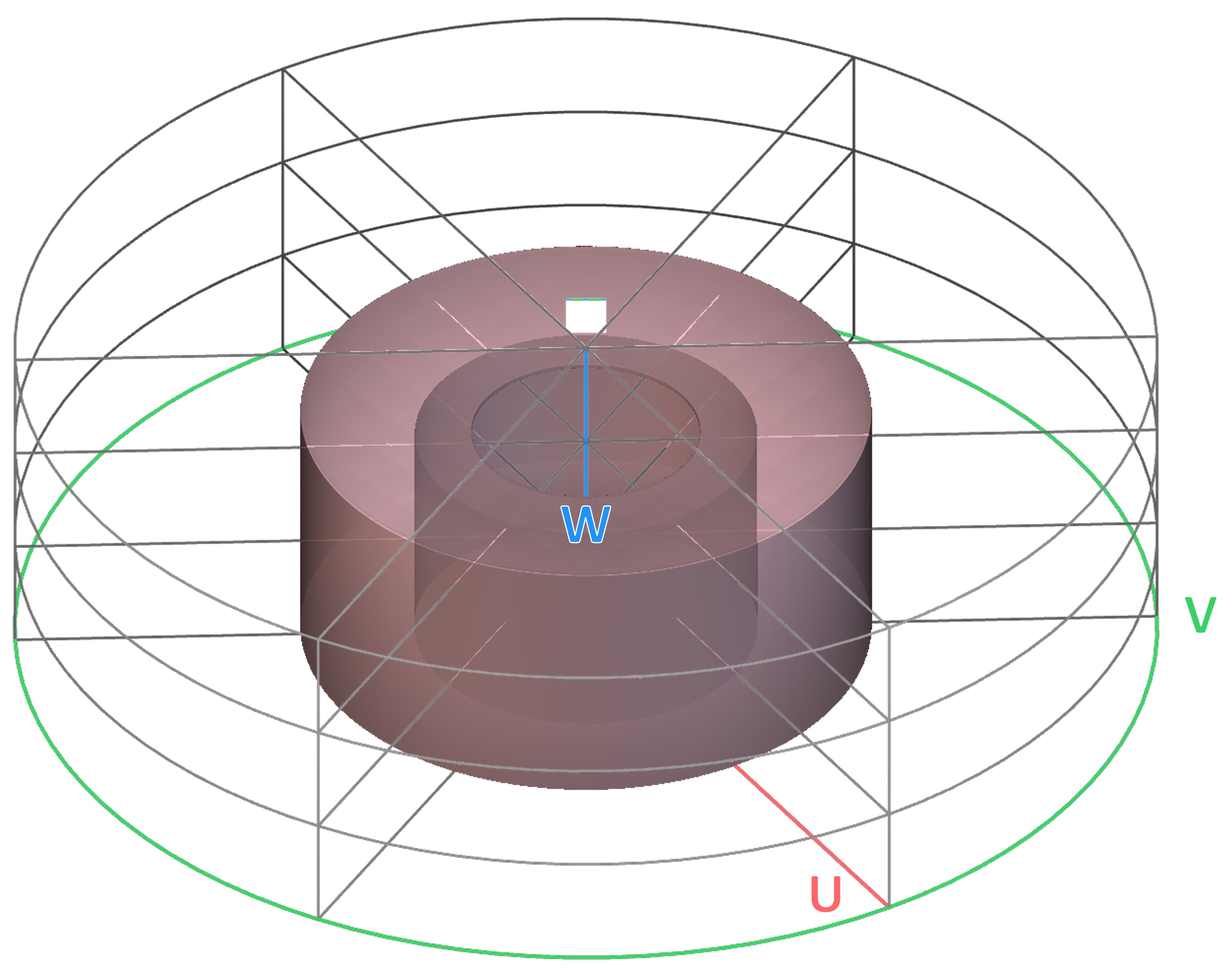}
    \caption{Cylindrical cell map with volume for lattice (red) and cell dimensions, in the form U x W x V, of 10 x 2 x 6}
    \label{fig:19}
\end{figure}

\begin{figure}
    \centering
    \includegraphics[width=1\linewidth]{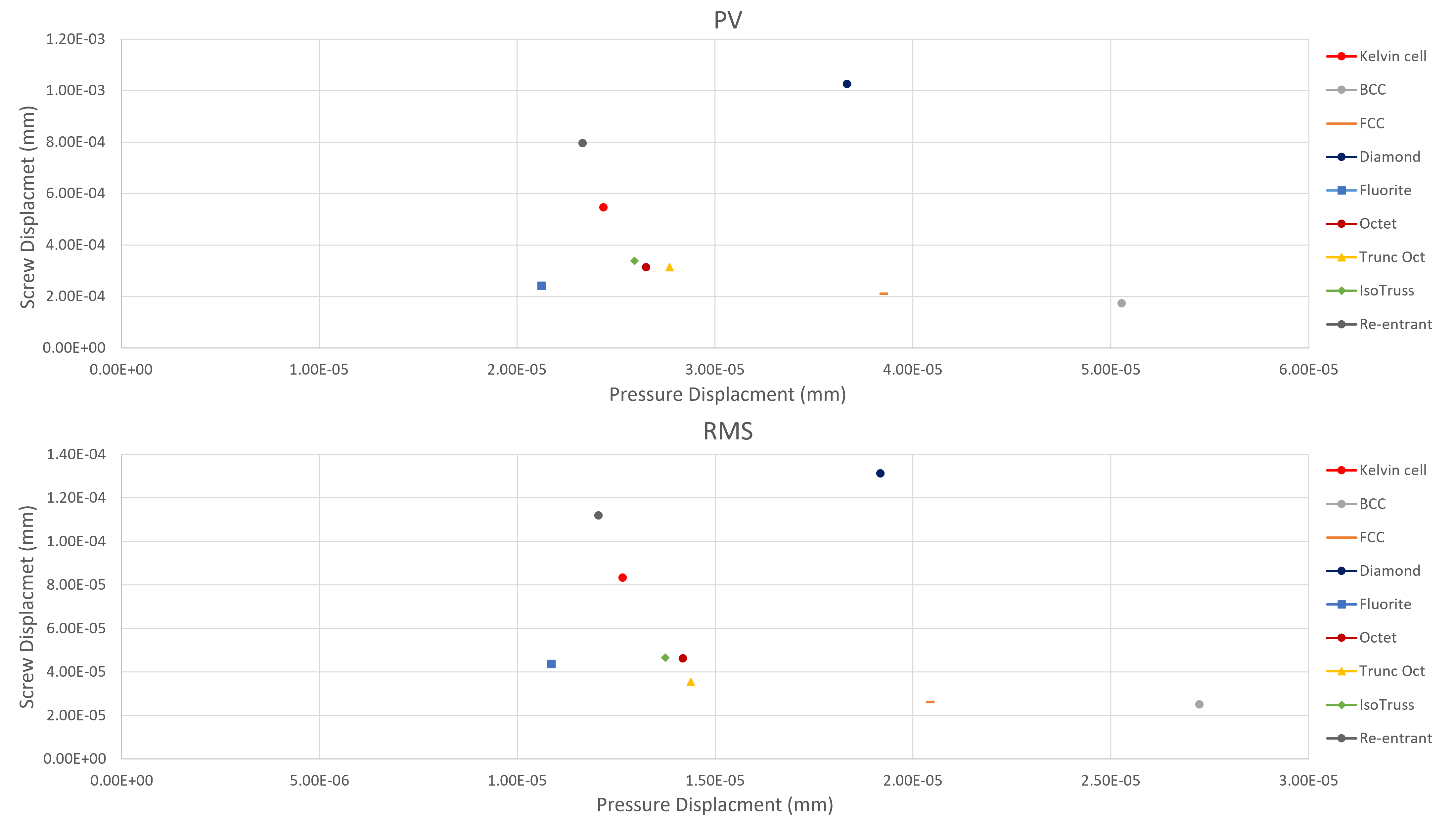}
    \caption{Results from initial mounting lattice test}
    \label{fig:18}
\end{figure}

The volume for the lattice was kept very small, this was for several reasons. Firstly these were added features, they would be adding mass and not reducing it, so this effect was minimised. The second reason was to try and keep the overhanging areas of the lattice to a minimum, this is a requirement of the printing process and will be discussed in Section \ref{sec:5}. The small volume and want for flexibility led to graph unit cells being tested. A cylindrical cell map was also selected due to the shape of the volume, seen in  Figure \ref{fig:19}.
In the initial test, the top surface displacement data was collected in the same way as it was in Section \ref{sec:Internal Lattice Research} and nine graph unit cells were compared on the PV and RMS values of the screwing load versus the pressure load, as shown in the results in Figure \ref{fig:18}.

\begin{figure}
    \centering
    \includegraphics[width=0.75\linewidth]{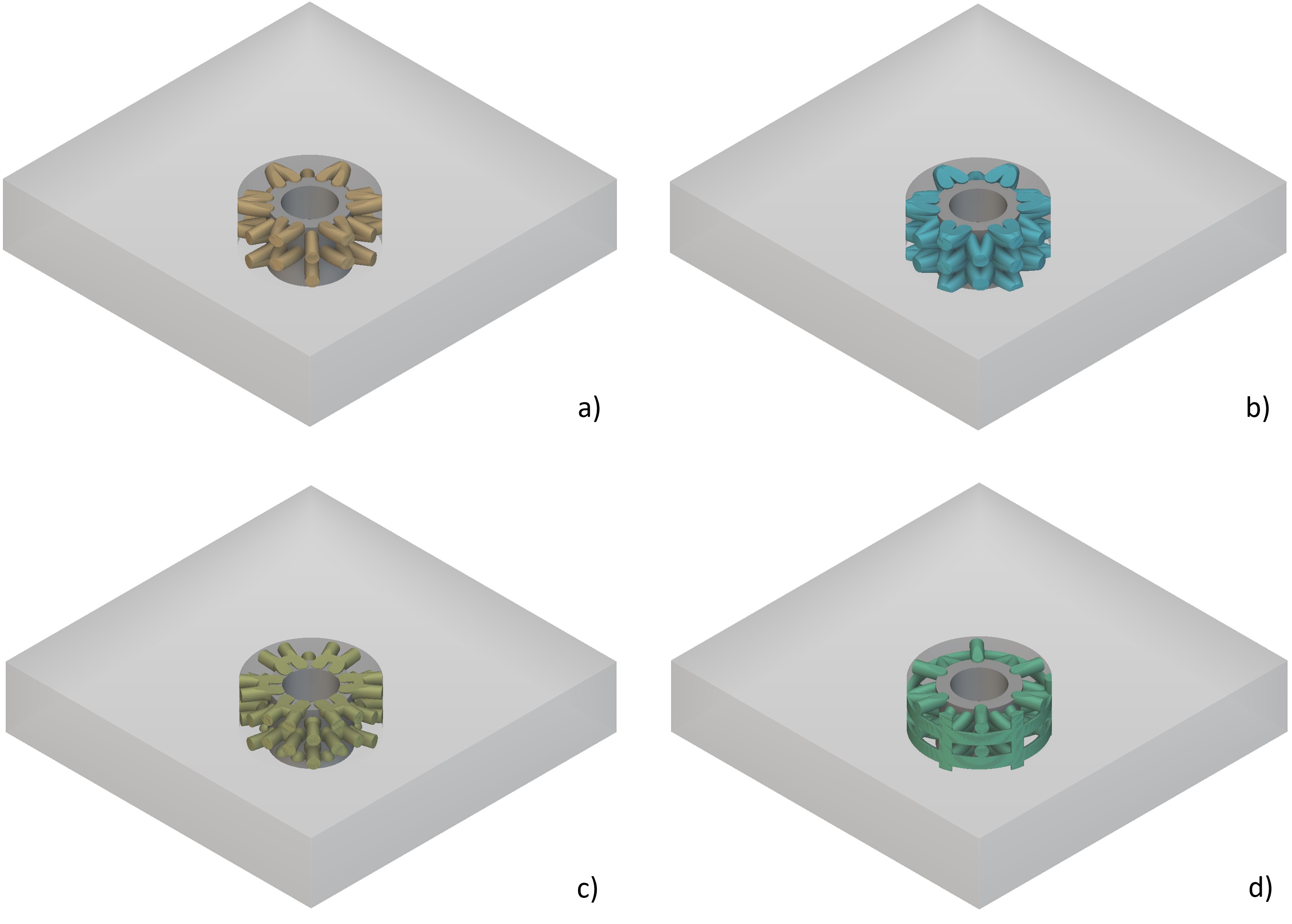}
    \caption{Four best-performing unit cell types: \textbf{a)} Face Centred Cubic lattice, \textbf{b)} Fluorite lattice, \textbf{c)} Truncated Octahedron lattice, \textbf{d)} IsoTruss lattice}
    \label{fig:20}
\end{figure}

From these results, four lattices were selected for further tests; Face Cubic Centre (FCC), Fluorite, Truncated Octahedron and IsoTruss, shown in Figure  \ref{fig:20}.  The octet lattice indicated good performance, but it was considered unsuitable for printing.

\subsubsection{Final Lattice Selection}
The shape of the four lattices was then explored, Figure \ref{fig:19} shows the cylindrical cell map and the region for the lattice (red). In the same way that the dimensions X, Y and Z could change in Figure \ref{fig:6}, U, W and V can as well, varying the cell size in the radial, height and the number of cells in the circumferential direction respectively. 

\begin{figure}
    \centering
    \includegraphics[width=0.75\linewidth]{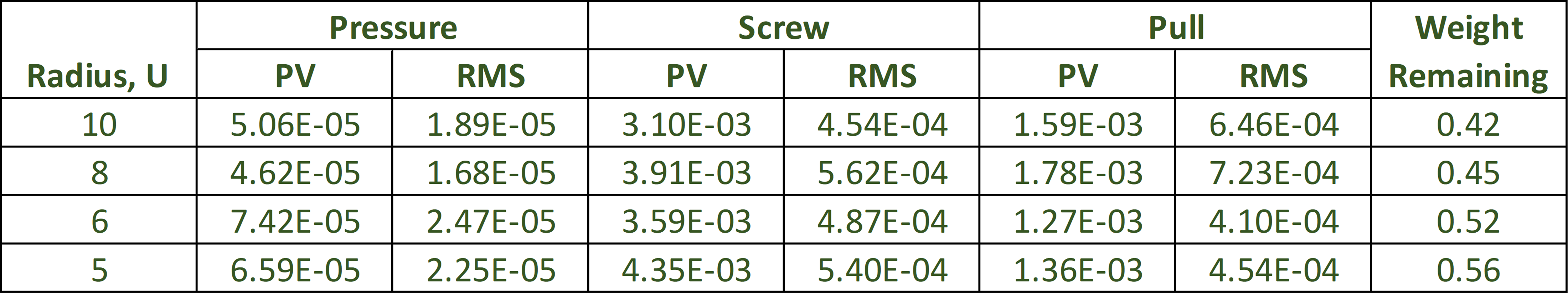}
    \caption{Table of IsoTruss experiment results}
    \label{table 2}
\end{figure}

While printability was a factor in the internal lattice selection, it had a larger impact on the mounting features due to the small design volume present. There also had to be enough room between the lattice struts for the powder to escape, especially as there was an elliptical cavity at the top that would have some unsintered powder in it after printing. As such, W was set to \SI{2}{\mm} and did not change as any more would have made the lattice too dense. The number of cells in the circumference, V was only slightly varied between 10 and 4 with 6 resulting in the optimal amount to balance between support and density. This only left the radius, U, to change.  
For each of the four types of unit cells, with W and V set as above, U was changed to 10, 8, 6 and 5. The PV and RMS displacement of the top surface was then measured for each of the three load cases. The weight remaining was recorded also. Figure \ref{table 2} shows an example using the values for the IsoTruss experiments.

Any radius that had a weight remaining of more than 0.5 was then discounted as this was deemed too dense for powder removal and printability. The remaining values of U for each type were then combined and a matrix comparison was carried out. Each was ranked for the six measured displacements and the weight remaining. The best performing overall was the IsoTruss lattice with a radius of \SI{10}{\mm}, therefore this one was selected for the design.

\subsubsection{Incorporation with Main Design Body}
\begin{figure}
    \centering
    \includegraphics[width=0.75\linewidth]{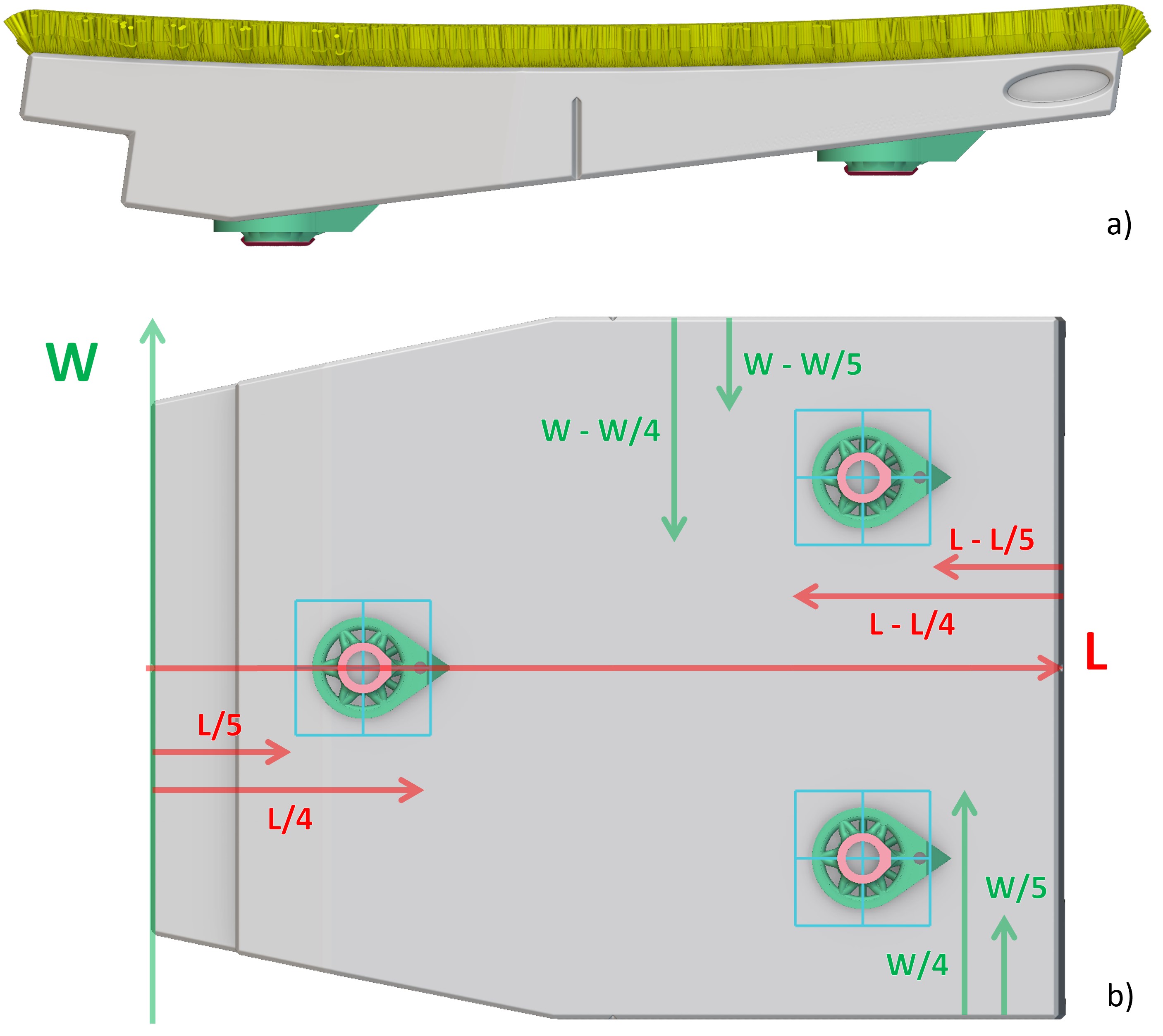}
    \caption{Experimental set-up for determining the optimal location of mounting points: \textbf{a)} Boundary conditions used, pressure applied to the mirror surface (yellow) and fixed at mounting points (red), \textbf{b)} Mounting point position system}
    \label{fig:21}
\end{figure}

\begin{figure}
    \centering
    \includegraphics[width=0.75\linewidth]{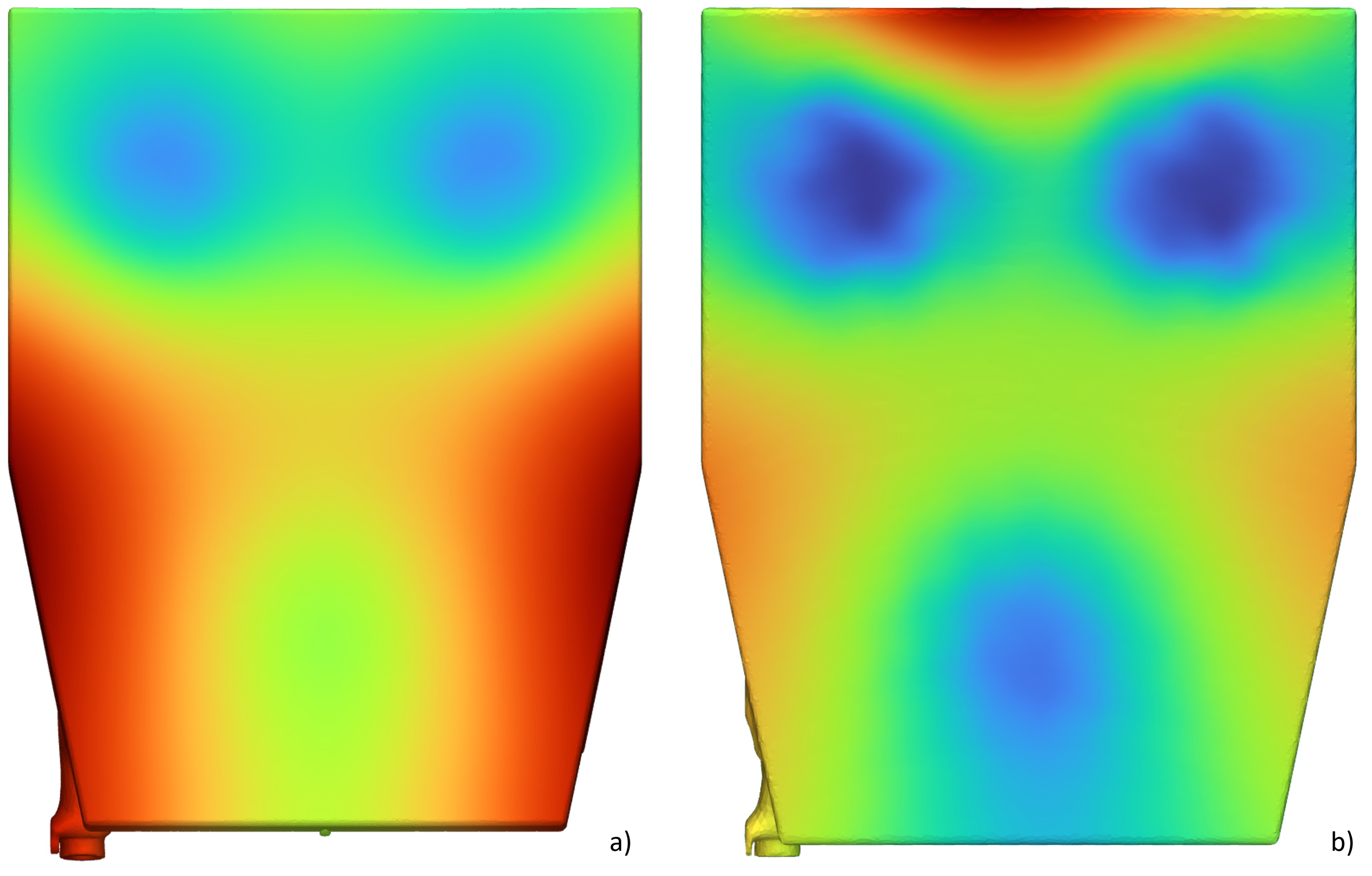}
    \caption{FEA displacement results for determining SPDT mount positioning: \textbf{a)} Closed-back design, \textbf{b)} Open-back design}
    \label{fig:22}
\end{figure}

An optimised design of the mount was then finalised for use on the main design, with an outer cylinder replacing the outer box and further features added to support printability (Section \ref{sec:Designing Against Overhangs}). It was decided that there should be three mounting points, as is used in most optical manufacture to prevent over-constraining the part.. 
The location of each mount was then considered, the aim was to best support the mirror and minimise deflection of the surface. A brief FEA experiment took place that trialled multiple positions on the design. Both the closed-back and open-back designs were orientated with the centre of curvature (CoC) of the mirror at the lowest point of the top surface, as this is the orientation they will undergo SPDT. A pressure force was then applied to the top surface and the mirror was fixed at the three mounting points. It was quickly determined from some initial simulations that, due to the part being thinner at the back, having two points at the back and one at the front was much more optimal. Figure \ref{fig:21} shows the boundary conditions and the three points defined by distances L and W. The range of distances is shown as well, between 0.2 and 0.25  times the distance from the edge. This range was chosen as it is the recommended range of support points for a simply supported beam and encompasses Airy Stress and Bessel points. 
The experiment concluded that a distance of 0.23L and 0.23W would be most optimal, the reason for this can be seen in Figure \ref{fig:22} which shows the FEA displacement results for the open-back and closed-back designs. There was a drive to keep the position the same on both designs to aid in the machining later, so 0.23 was selected due to it supporting both designs equally, not supporting the weaker area of one over the other.

\subsection{Topology Optimisation of the Nanosat Interface Features}
\label{sec:Topology Optimisation of the Nanosat Interface Features}
\begin{figure}
    \centering
    \includegraphics[width=0.75\linewidth]{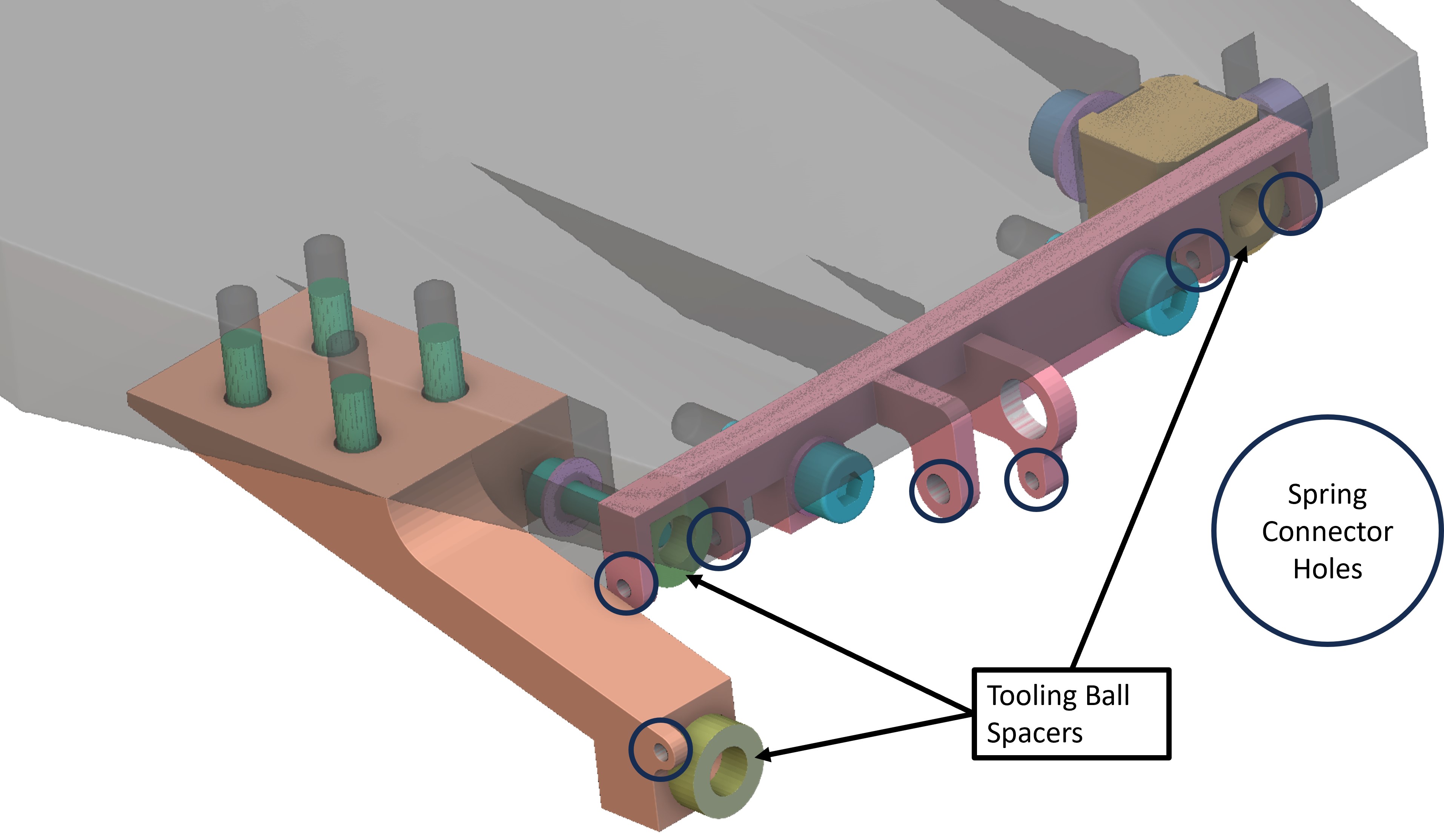}
    \caption{Original interface with 18 parts shown through colour and fixed features highlighted.}
    \label{fig:23}
\end{figure}

Besides the main body, there are 18 other parts that make up the interface that connects the mirror to the nanosat. The new AM mirror must accept the same three tooling balls and seven springs that locate the mirror for the deployment mechanism. The dimensions and locations of these holes and features must therefore remain the same to ensure the new design would function the same as the original. Figure \ref{fig:23} shows the 18 parts that made up the original interface as well as the ten fixed features that must not change.
This entire interface was designed for subtractive manufacture, hence the need for washers, grooves and screws to hold various parts in place. However, when re-designing this part for AM, most of them could be removed and only the parts that housed one of the fixed features needed to be kept. This was mainly two of the features, the spring connector interface (red) and the M3 arm (orange).

\subsubsection{Spring Connector Interface Topology Optimisation}
\begin{figure}
    \centering
    \includegraphics[width=0.75\linewidth]{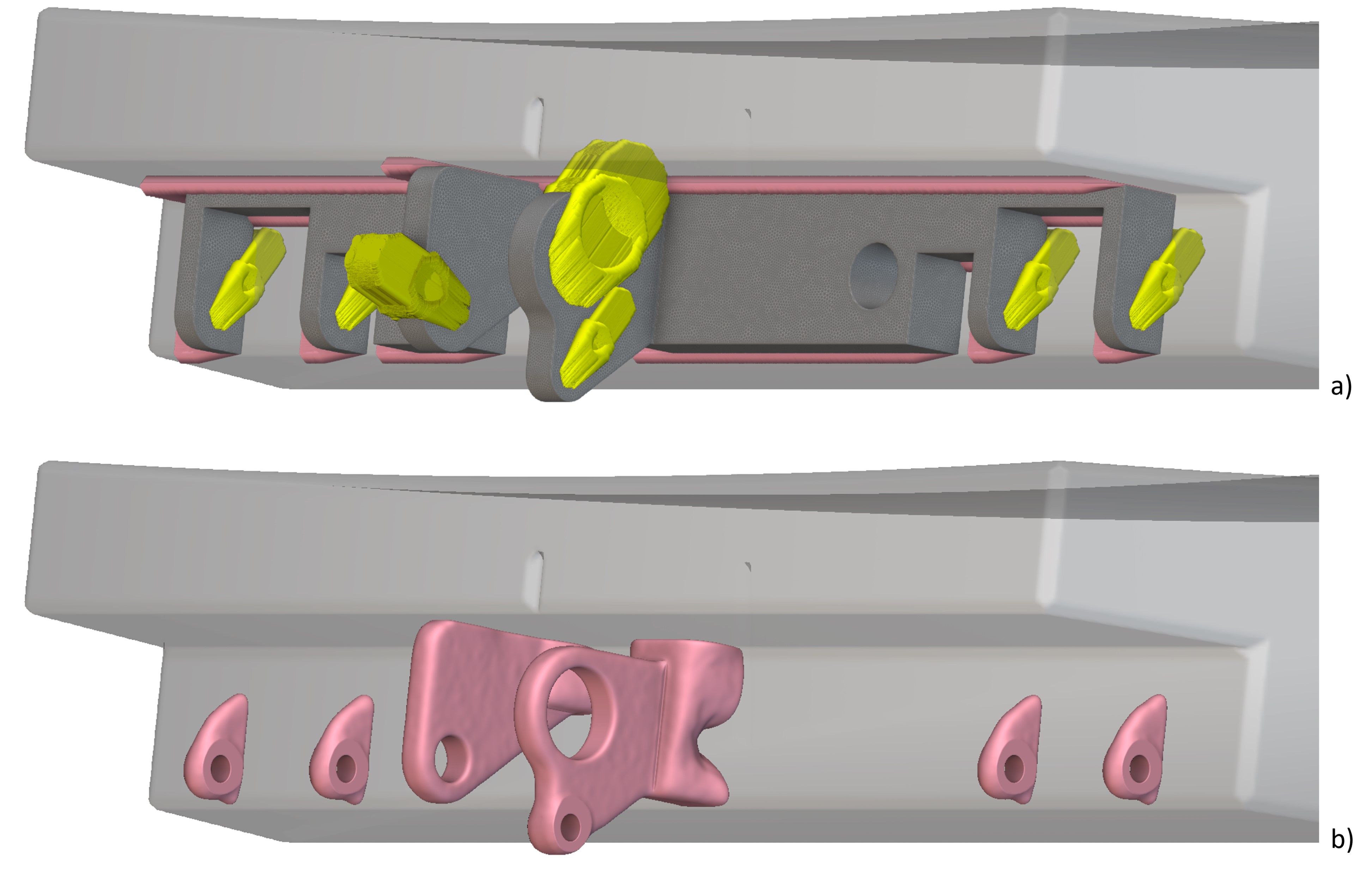}
    \caption{Topology optimisation of spring connector interface: \textbf{a)} Design volume with spring vector forces (yellow) and fixed rear and upper face (red), \textbf{b)} Topology optimised result at 0.25 volume remaining}
    \label{fig:24}
\end{figure}

Two of the tooling ball spacers were simply joined to the main body as there was no benefit in optimisation. The large piece in the centre of the interface (seen in red on Figure \ref{fig:23}), was the first to be optimised. Figure  \ref{fig:24}a shows the boundary conditions that were set up for the topology optimisation, where the back and the top face of the part are fixed and the vector forces representing the springs were applied. A small amount of material was added in this case to the underside of the two central features, this added a 45° chamfer and removed an overhang which was present when the part was in its build orientation, which is discussed in Section \ref{sec:Designing Against Overhangs}. The optimisation objective was a structural compliance response and the optimisation constraint was volume fraction, which meant that material that experienced the most stress was prioritised and a volume reduction of 0.25 was the aim. The resultant topology is shown in Figure  \ref{fig:24}b.

\subsubsection{M3 Arm Topology Optimisation}
\begin{figure}
    \centering
    \includegraphics[width=0.75\linewidth]{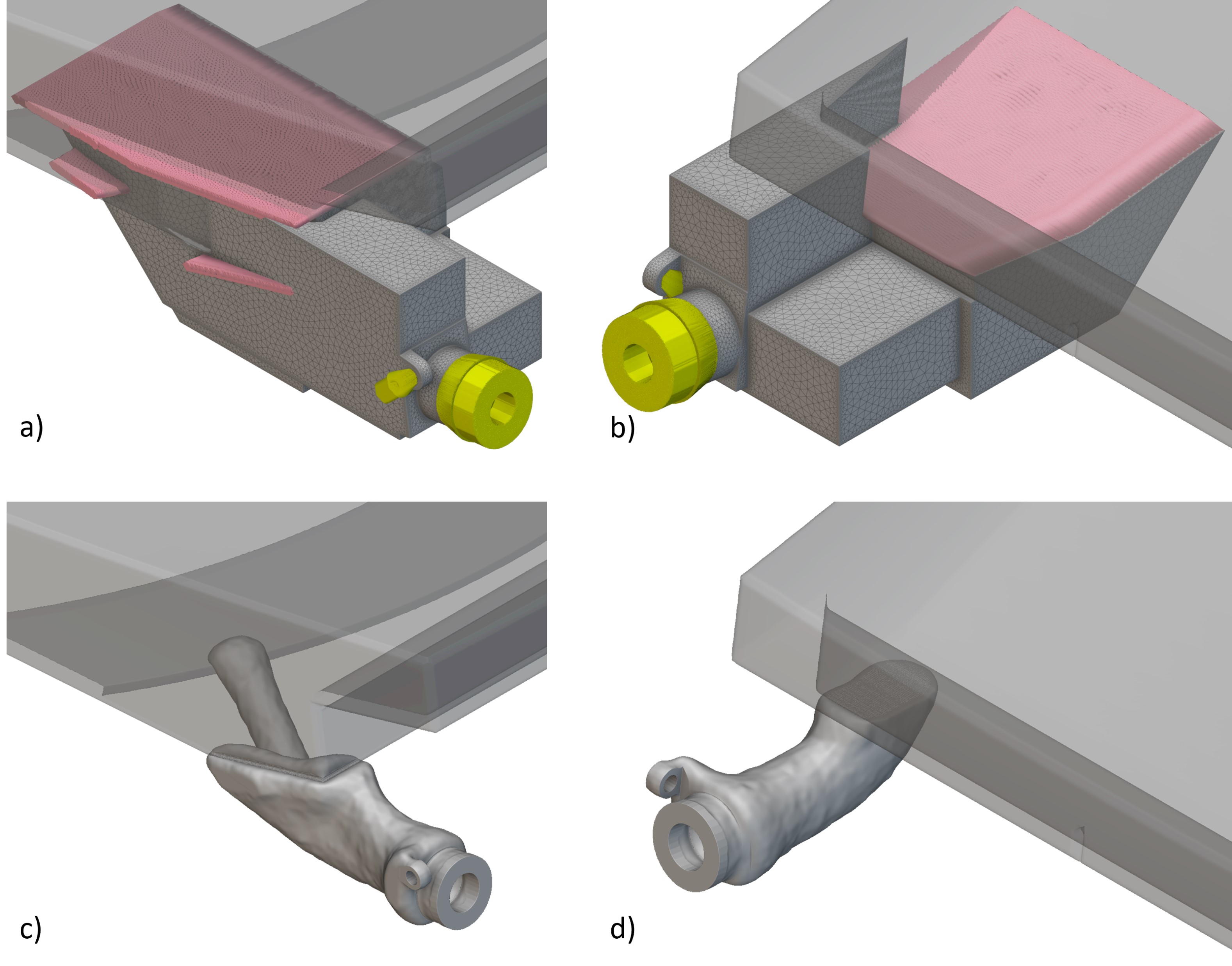}
    \caption{Topology Optimisation of M3 arm: \textbf{a)} Design volume of open-back design with spring vector forces (yellow) and three fixed surfaces on the back (red), \textbf{b)} Design volume of closed-back design with spring vector forces (yellow) and the fixed surface on the back (red), \textbf{c)} Topology optimised result for open-back design at 0.1 volume remaining, \textbf{d)} Topology optimised result for closed-back design at 0.1 volume remaining}
    \label{fig:25}
\end{figure}

The other feature selected for topology optimisation was the M3 arm (shown in orange in Figure \ref{fig:23} ). This part was more separated from the body and allowed for more experimentation. The main force acting on this part was a reactionary force from the tooling ball pushing into the flexure on the nanosat, and while the original design provided sufficient support, its geometry, like the rest of the original design, was dictated by its ability to be machined. Therefore a larger design volume was used on this feature, being careful not to expand into the fixed features or the design envelope of the whole mirror. This design volume, along with the smaller force of the spring, can be seen in Figure  \ref{fig:25}a. The fixed area in this case was a lot larger and less restrictive, as the whole of the back of the mirror could be used. As mentioned in Section \ref{sec:Internal Lattice Research} there will be two different back designs, which means two slightly different surfaces to interface with, resulting in two different topology-optimised M3 arms. The parameters for the optimisation are the same as the spring connector part, with a different volume reduction constraint of 0.1 used to accommodate the extra design volume added. The open and closed back topology optimised M3 arms are shown in Figure  \ref{fig:25}b.

\section{Features Included to Aid Printing and Post-Machining}
\label{sec:5}

Besides the three major objectives, discussed in the previous sections, numerous smaller features needed to be added to make the part printable and to aid in the machining stages afterwards. 

\subsection{Powder Removal}
\begin{figure}
    \centering
    \includegraphics[width=0.75\linewidth]{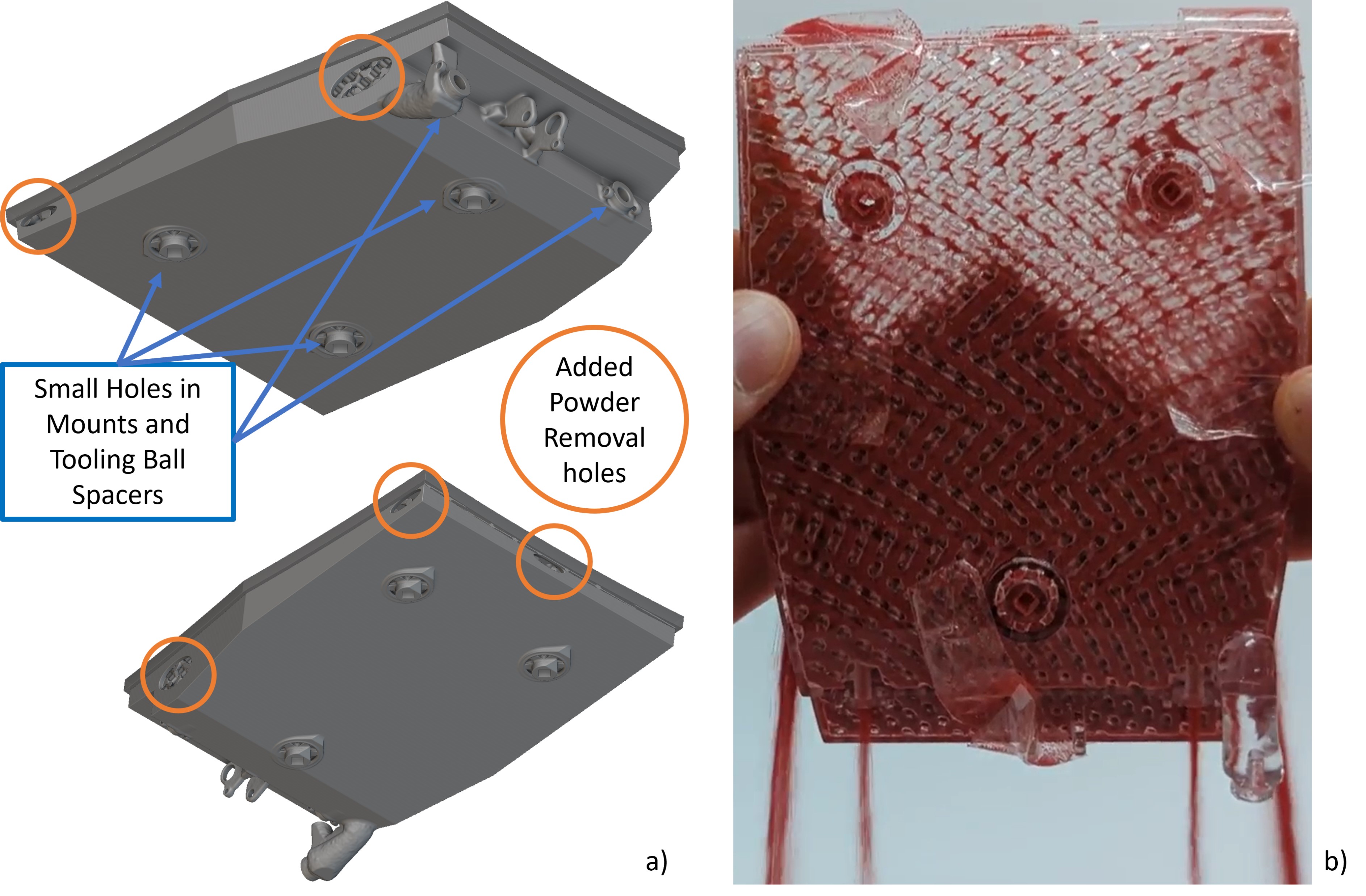}
    \caption{Powder removal model and test: \textbf{a)} Model used for powder removal tests with existing and added holes, \textbf{b)} Powder removal test with all holes open}
    \label{fig:26}
\end{figure}

Parts to be printed with L-PBF have to consider how the unsintered powder will escape the part during extraction, especially with complex geometries like lattices. Both lattices were selected with this in mind and the fairly low density aided this consideration. The open-back design was not considered a concern for powder removal; precautionary visual checks were enough to validate that powder could easily escape from every crevice. The closed-back design, however, would have lots of powder trapped between the internal lattice, so holes would be needed in the outer shell. To decide the location, number and size of these holes a design was printed in ClearVue resin on a stereolithography machine. This material was picked to give a dimensionally accurate part that is also transparent. A very simple powder removal test then took place, the top surface was printed separately so it could be removed like a lid, allowing red sand to be poured in to act as the unsintered metal powder. The lid was then fixed and each of the holes on the side were uncovered sequentially in a series of tests, from smallest to largest. Figure  \ref{fig:26}a shows the models printed with each of the holes that could be used for powder removal, and Figure  \ref{fig:26}b shows the test being carried out. 
It was hoped that the tooling ball holes and holes in the mounting feet would be sufficient for powder removal, however, the decision was taken to add the two small holes at the back corners of the design, as these performed very well at allowing the powder to escape and it was hoped that locating them near the corners would maintain the shells strength. This decision was taken due to uncertainty in the quality of the other holes and how well they would be printed, or in fact if the holes would be used in this final design at all and would be instead be drilled in afterwards.

\subsection{Designing Against Overhangs}
\label{sec:Designing Against Overhangs}
\begin{figure}
    \centering
    \includegraphics[width=0.5\linewidth]{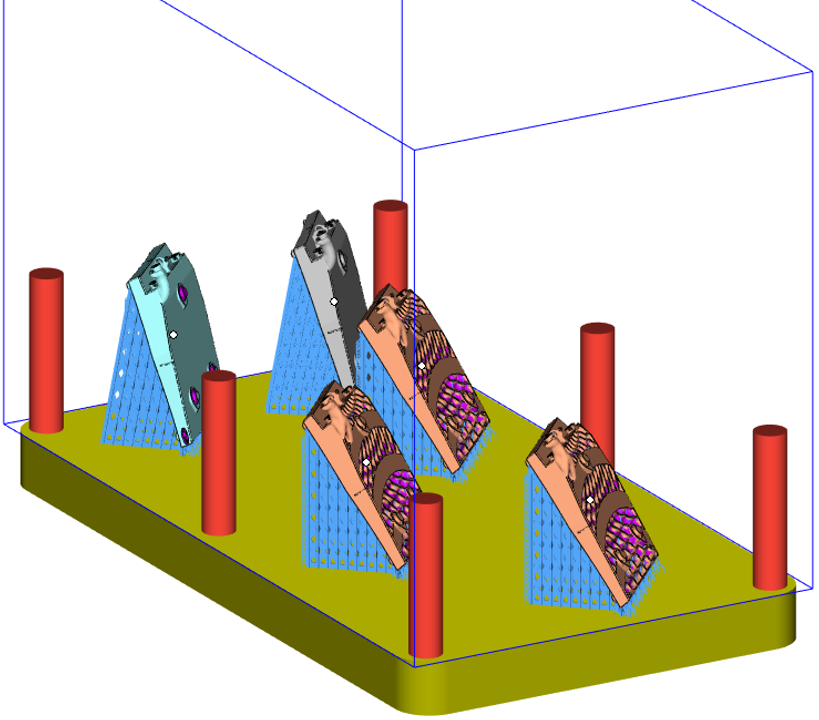}
    \caption{Mirrors orientated on the build plate with generated support shown (blue). Image credit: Martin Sanchez, CA Models }
    \label{fig:27}
\end{figure}

The orientation of the part during printing has a large impact on many aspects of the design; overhanging regions that cannot be supported are the main concern from a design point of view. Some features can be augmented to mitigate them, but others, such as the main mirror body, cannot. After conversations with the printing engineer, the orientation decided to best accommodate these features were the orientations seen in Figure  \ref{fig:27}, with the angle between the normal to the CoC and the build plate between 0° and 45°.
Other features could then be designed to accommodate this angle, such as the features added to the spring connector interface mentioned in Section \ref{sec:4.}. The mounting points (Section \ref{sec:Designing Mounting Points for SPDT}) were designed for this also, they protruded out from the main body and so a 45° taper feature was added to two designs for the open-back and closed-back mirrors. The printing engineers also had some concerns about regions of the lattice on the open-back design. There were areas on the back surface where the lattice ended that did not connect, this left thin protruding geometries that risked overmelting due to a lack of heat dissipation or potentially curling up under the heat. If the re-coater blade of the printer hit these geometries it would result in a failed print. Therefore to address this, ribs were added to the back of the open-back design. This not only removed these unwanted features but also added some structural support and aided the printing of the mounting points.

\subsection{Features Added to Aid in Post-Printing Machining}
\label{sec:Features Added to Aid in Post-Printing Machining}
\begin{figure}
    \centering
    \includegraphics[width=0.75\linewidth]{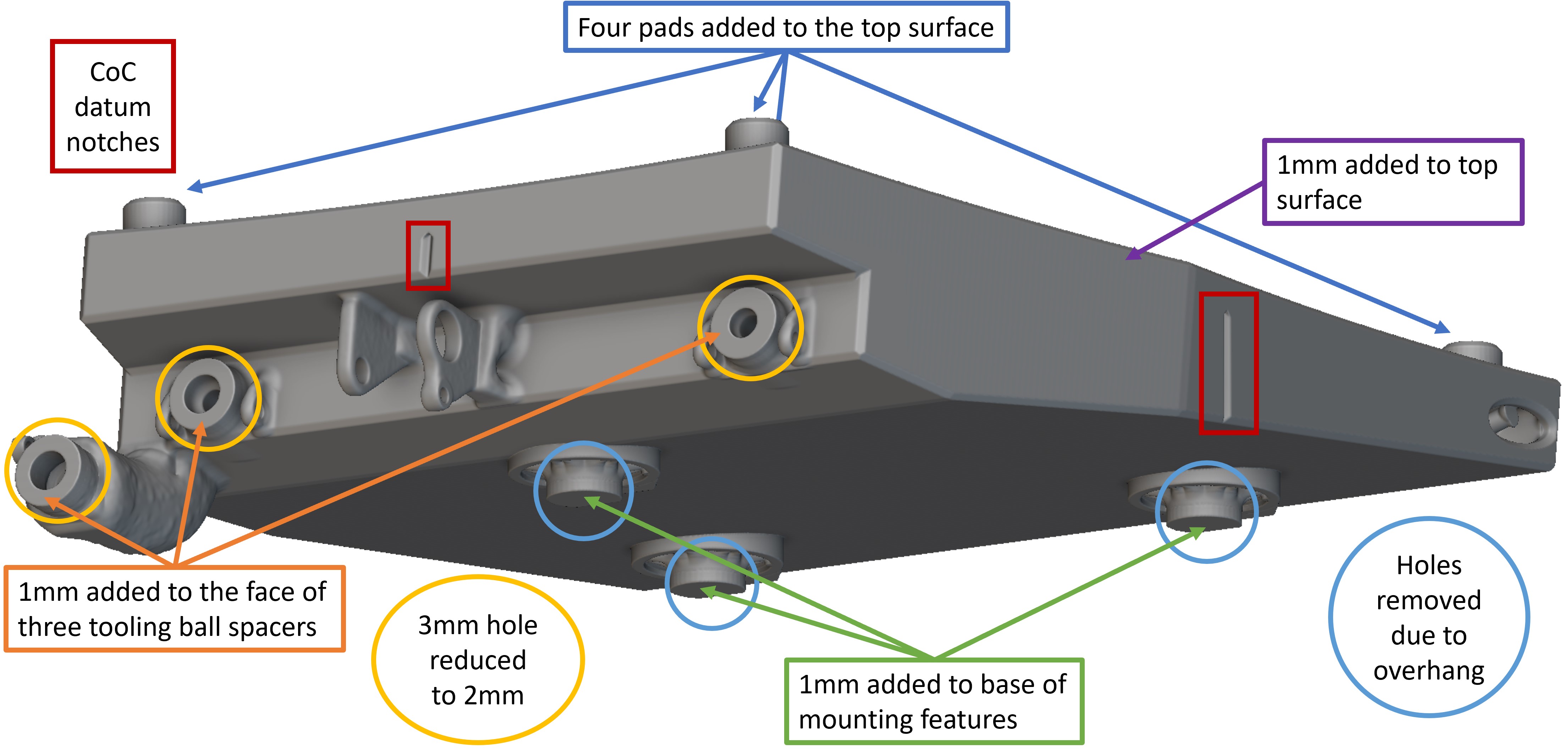}
    \caption{Features altered for machining}
    \label{fig:28}
\end{figure}

Numerous features on the design were added to aid in the machining of the printed part. It can often be complex to accurately machine an AM part, largely due to the lack of precision in the results from L-PBF. Machining requires designated datum planes to be established so that other features can be dimensionally referenced from that plane. In subtractive machining, this often takes the shape of an initial surface that is machined, however in AM, those surfaces are already established, therefore one of those has to act as a datum for machining operation. The relatively low dimensional accuracy, however, makes this challenging. The approach of this study was to include four sacrificial pads to the optical surface that could be machined flat to form a datum surface. A similar approach was taken in the paper by \textit{Westsik, M., et al. (2023)} \cite{Marcell}, however, with some stability concerns occurring during machining, the decision was made to include four pads here. A notch on each of the four sides, in line with the CoC, would then provide the other two planes. Along with some added physical measurements, this would then give enough information to machine the other required areas. These areas include high-tolerance surfaces and holes that must interface with something external, such as tooling balls or SPDT jigs. All these features are highlighted in Figure  \ref{fig:28}

\subsection{Final Design}
\begin{figure}
    \centering
    \includegraphics[width=1\linewidth]{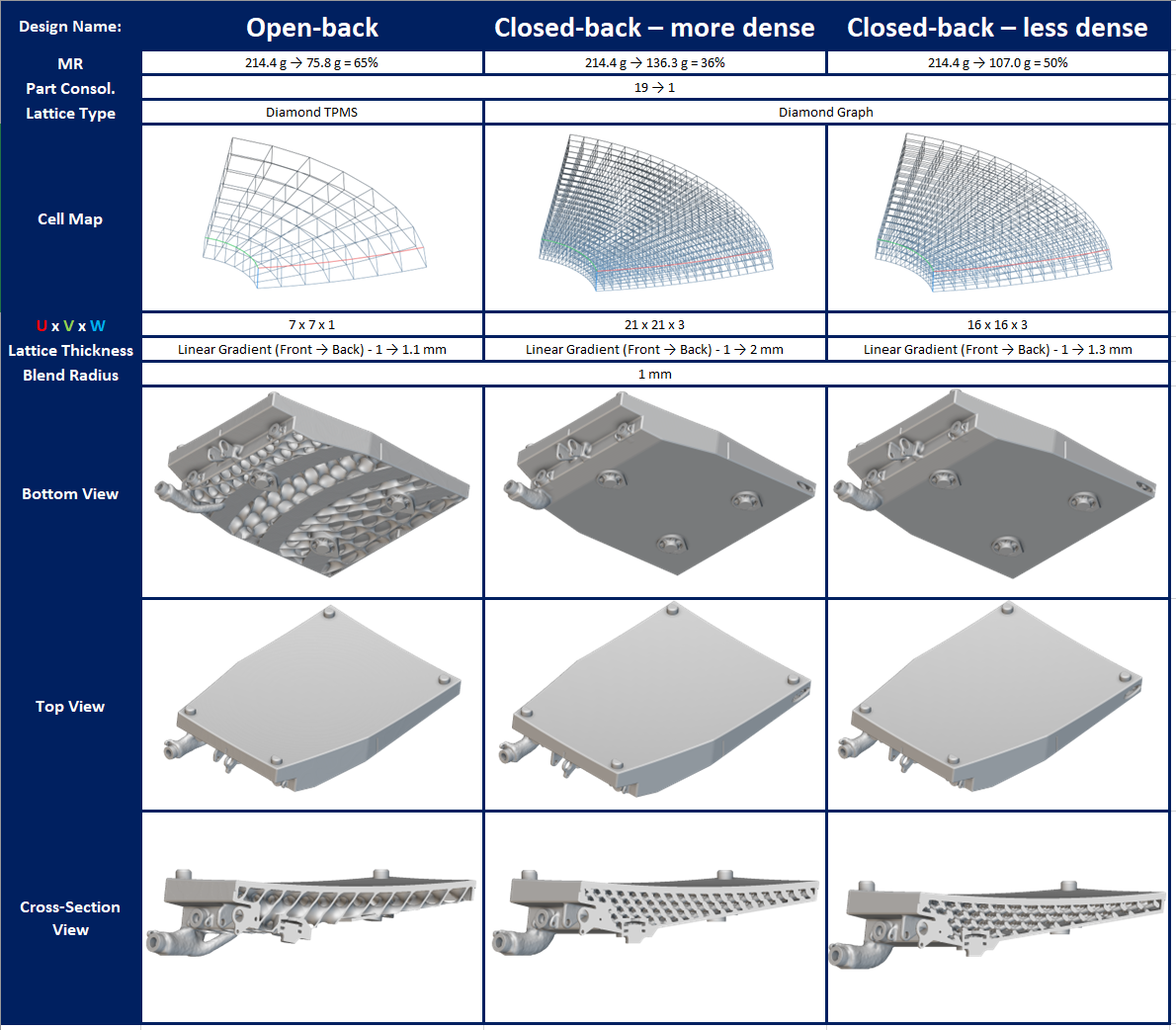}
    \caption{Table of parameters for each of the three designs that were printed}
    \label{table 3}
\end{figure}

The series of design choices outlined in the previous two sections, culminated in two designs, one with an open-back and one with a closed-back.
The mass of each design was then simulated and the overall MR of each design was compared to the original. With all of the features added and slight shape changes, the MR for the open-back was 65\% and the closed-back was 36\%. This was slightly lower than desired and therefore the internal lattice was investigated to see if any modifications could be implemented. The open-back design already had a very stretched lattice and trying to make it any thinner resulted in a geometry that would not be printable, therefore no further changes made to this design. The closed-back design however, had much more flexibility in its lattice density, and so with the rest of the design features included, a less dense version of the design was established, which gave an overall MR of 50\%. However, with this less dense internal structure not having undergone as many tests as the more dense version the decision was taken to print one of each. With two different closed-back designs being printed, cost and process decisions, led to three of the open-back designs being printed. The final three designs that were selected for printing can be seen in Figure  \ref{table 3}, the final design parameters for the internal lattice along with part consolidation and MR numbers for each are shown as well.

\section{Printing \& Machining}
\label{sec:6}
\subsection{Printing}
\begin{figure}
    \centering
    \includegraphics[width=0.5\linewidth]{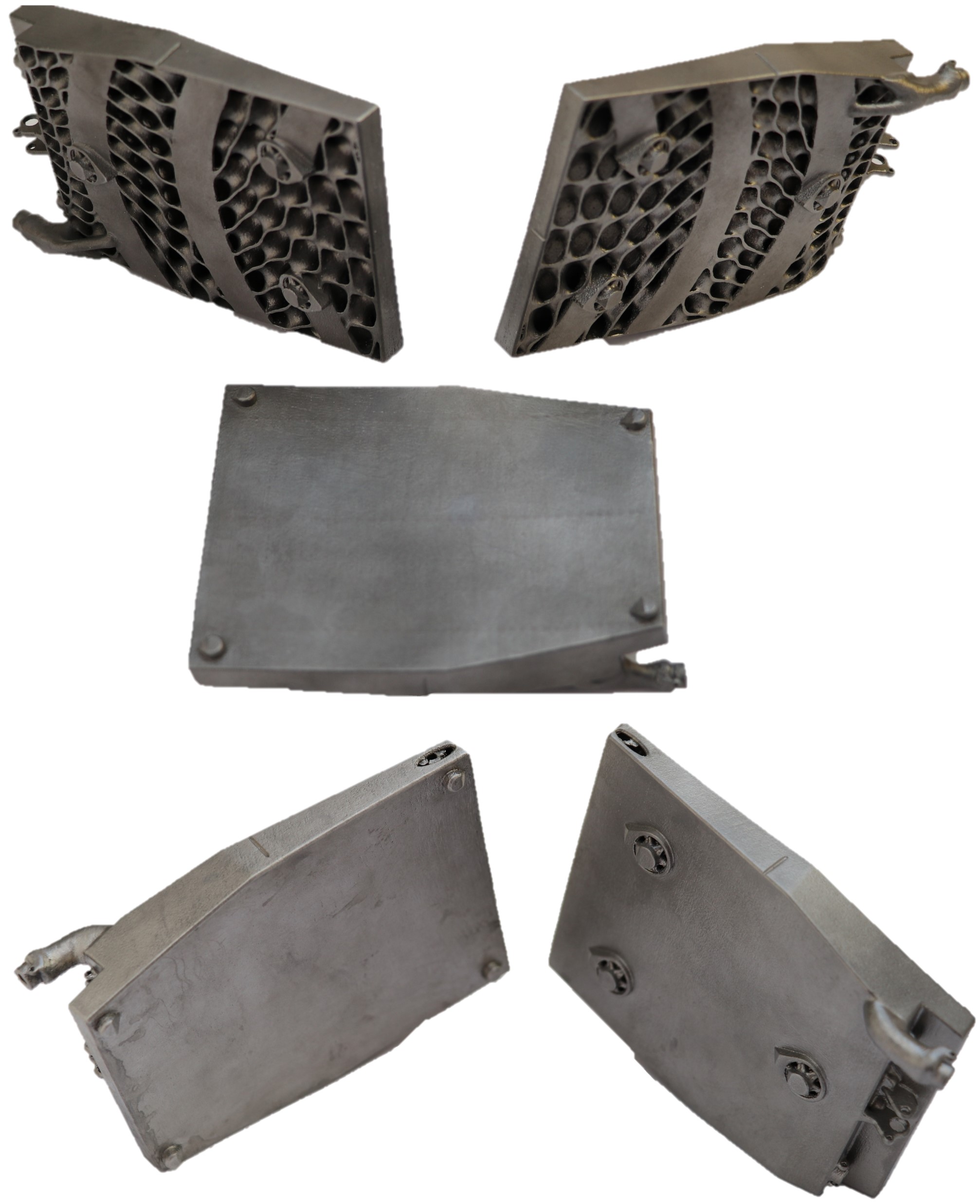}
    \caption{Five prints, after bead-blasting. Image credit: Jason Cowan  }
    \label{fig:29}
\end{figure}

\begin{figure}
    \centering
    \includegraphics[width=0.75\linewidth]{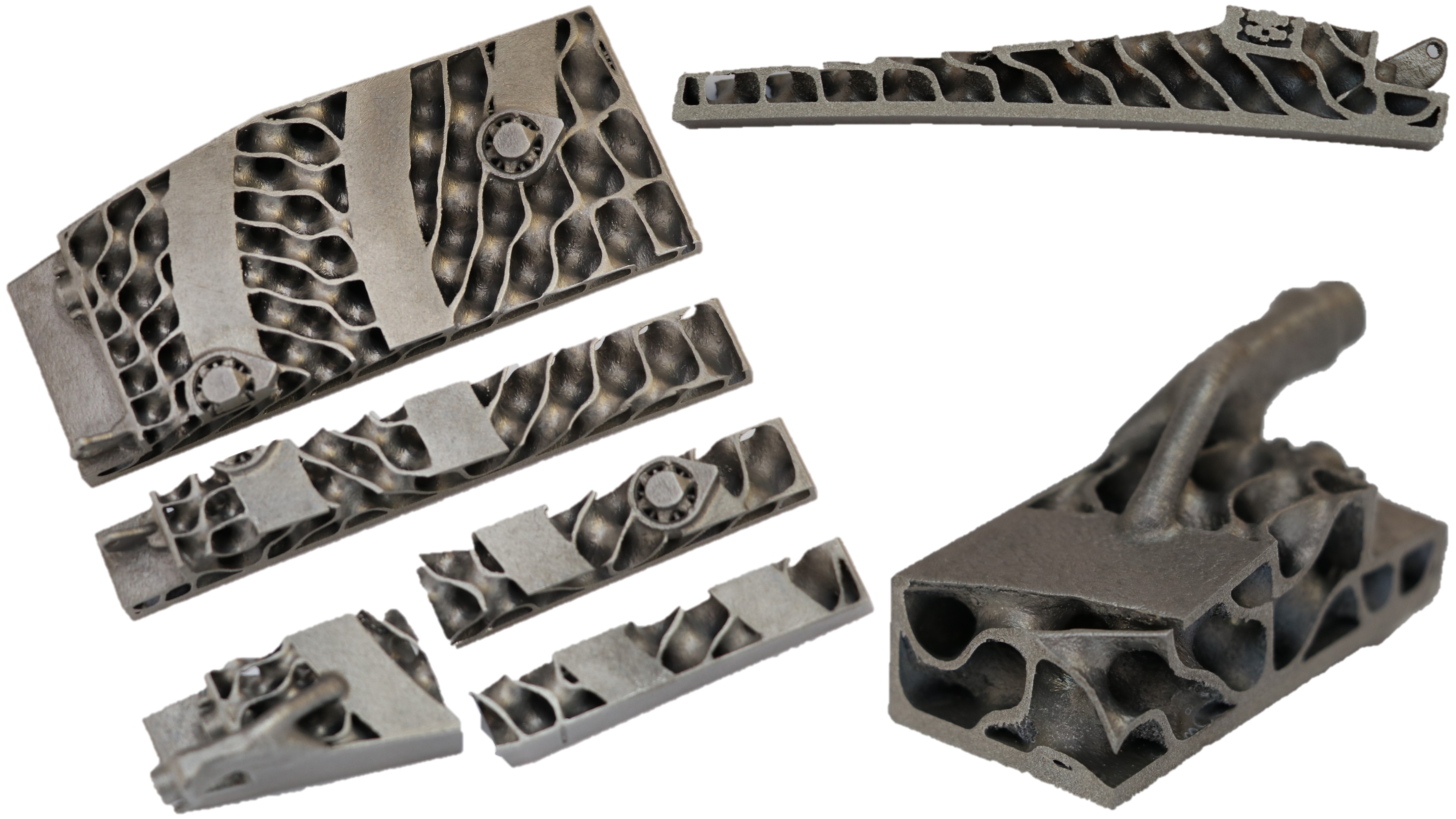}
    \caption{Wire - EDM open-back mirror. Image credit: Jason Cowan}
    \label{fig:30}
\end{figure}

Three open-back and two closed-back designs, one of each variant, were printed in AlSi10Mg on a L-PBF printer. Once removed from the build plate each print was then bead-blasted, a process which fires small glass beads at a surface at high speed, to clean and finish it. The printed parts can be seen in Figure  \ref{fig:29}. 
The quality of the prints looked good at inspection and all of the smaller features at the interface section seemed to have been accurately captured, with no trapped powder visible. To investigate the quality of the print further, one of the open-back designs was sectioned into smaller pieces using wire EDM (Electrical Discharge Machining), the resulting sections can be seen in Figure  \ref{fig:30}.
This process revealed a rougher surface finish on the underside of the lattice, this was as a result of the print orientation, with this region of the lattice being on the bottom and unsupported, leading to over sintering. The bead-blasting removed this finish from the outer surfaces but could not fully penetrate the lattice. The mounting features were also highlighted to have successfully printed to a good resolution as well, showing separation from the inner and outer cylinders. Two of these sectioned pieces will undergo X-ray Computed Tomography (XCT) to highlight design features such as the M3 arm and the mounting points.

Before the parts were sent for machining, each one was weighed. Each weighed slightly more than the computer models suggested. The open-back prototypes were 6.4\% heavier, and both designs of the closed-back prototypes were 1.3\% heavier. The increase in weight can perhaps be attributed to inaccuracies in the meshing of the files before printing or more powder being melted than expected by the laser.

\subsection{Machining}
\begin{figure}
    \centering
    \includegraphics[width=0.75\linewidth]{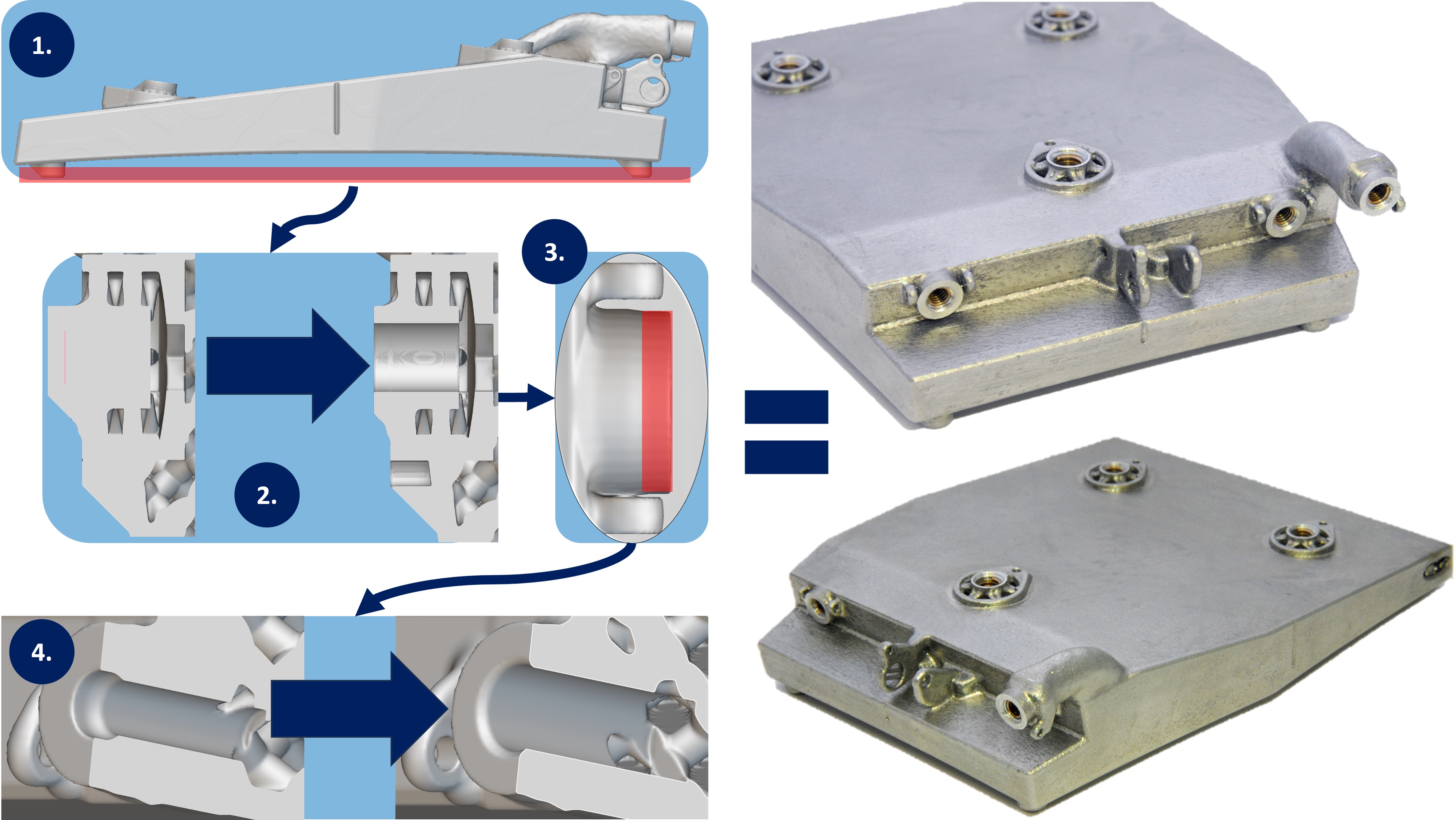}
    \caption{Machining stages preformed \& images taken before SPDT}
    \label{fig:31}
\end{figure}

To date, only the more dense, closed-back mirror has been machined fully, with the other two designs in progress. As such, the closed-back, more dense design, will be the one focused on in this section, with the process for the other designs being largely the same.  Figure  \ref{fig:31} shows the machining stages taken in preparing the mirror for turning, which correspond to the machining steps described below.

\begin{enumerate}
    \item  The top pads on the mirror face are machined flat to give a datum plane to set the part on
    \item Machining the mounts 
    \begin{enumerate}
        \item The bottom surface of the mount is machined flat
        \item A hole is drilled into the centre of the pad and then tapped with a thread, a helicoil is then inserted into the hole to provide added strength
        \item A dowel hole is drilled into the material below the main pad
    \end{enumerate}
\item The face of the tooling ball spacers on the spring interface are machined flat
\item The holes behind the spacers are drilled wider and then tapped with a thread
\end{enumerate}

\begin{figure}
    \centering
    \includegraphics[width=1\linewidth]{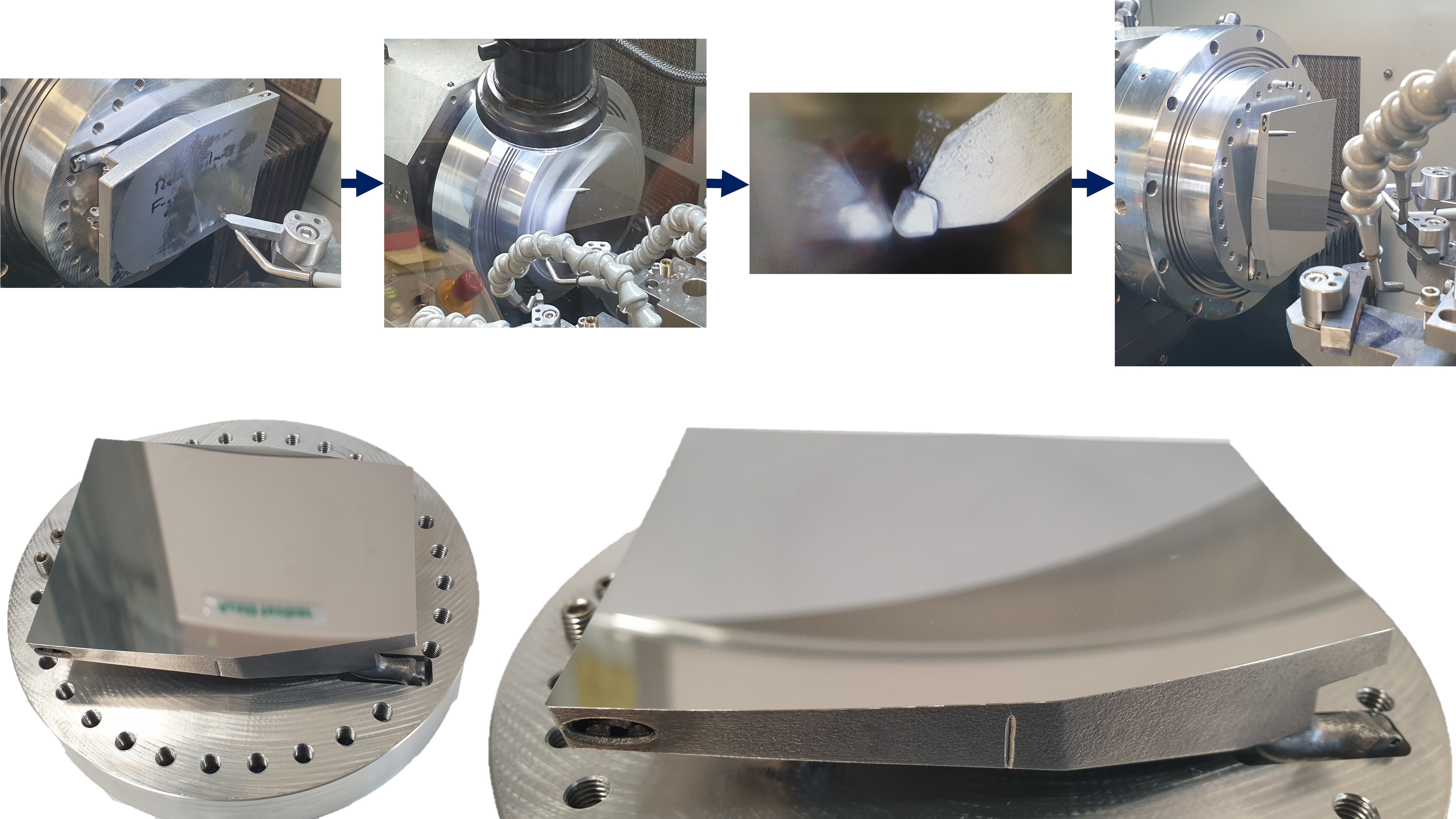}
    \caption{SPDT process and resulting mirror face}
    \label{fig:32}
\end{figure}

 The part could then be mounted to a custom-built jig, which would hold and orient the mirror with the CoC at the centre of rotation and the normal perpendicular to the jig surface. After a pass of CNC machining to remove the pads and roughly \SI{1}{\mm} from the top surface, the part was ready for SPDT. 
 The SPDT process is highlighted in Figure  \ref{fig:32}, showing the part mounted on the machine, the part being spun up, the cutter removing material and finally a reflective part, with the final mirror shown on the jig as well.

\section{Metrology}
\label{sec:7}
\begin{figure}
    \centering
    \includegraphics[width=1\linewidth]{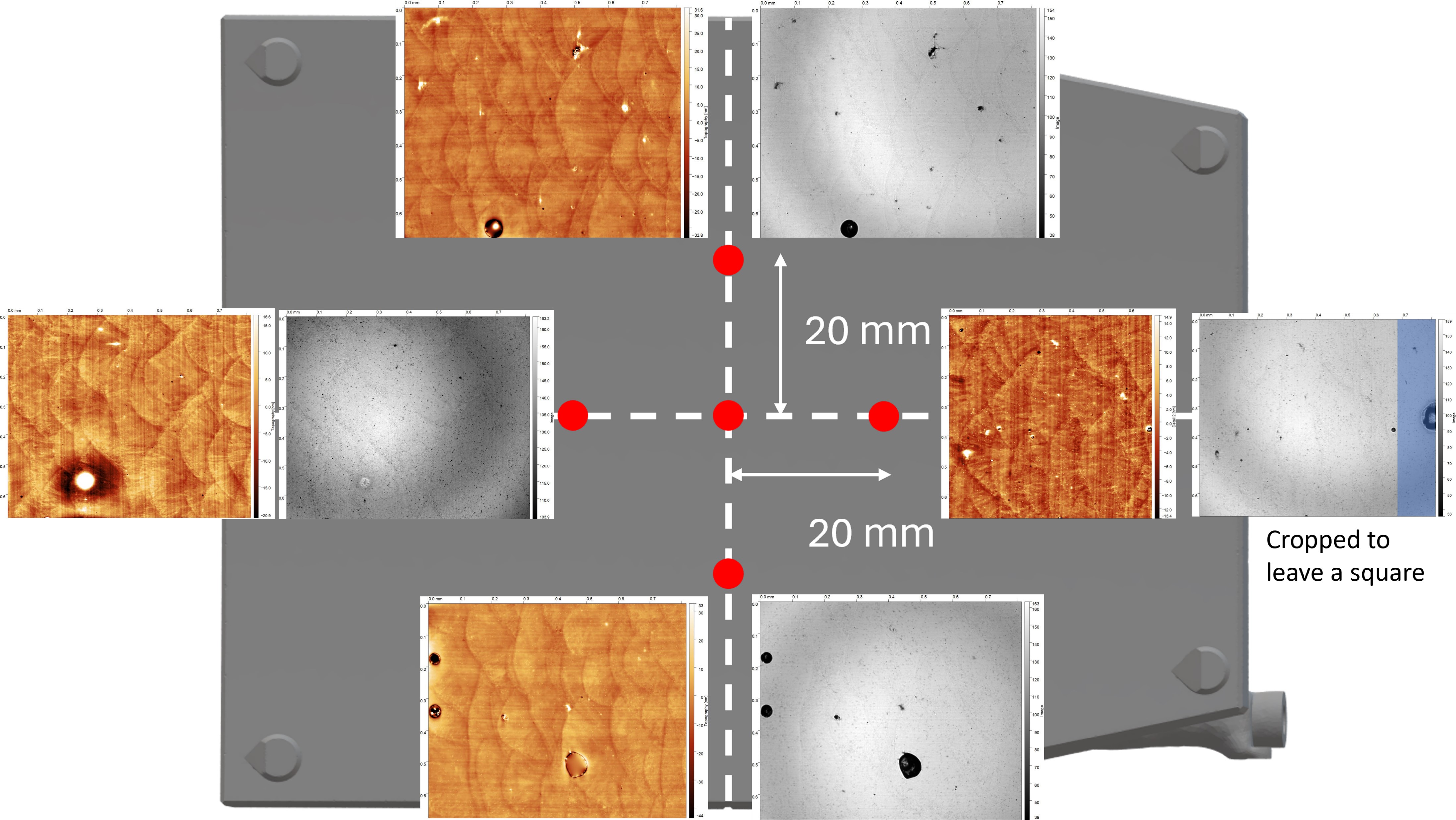}
    \caption{Micro-interferometry topography \textit{(left)} and microscope image \textit{(right)} at four sites on the more dense, closed-back design}
    \label{fig:33}
\end{figure}

\begin{figure}
    \centering
    \includegraphics[width=1\linewidth]{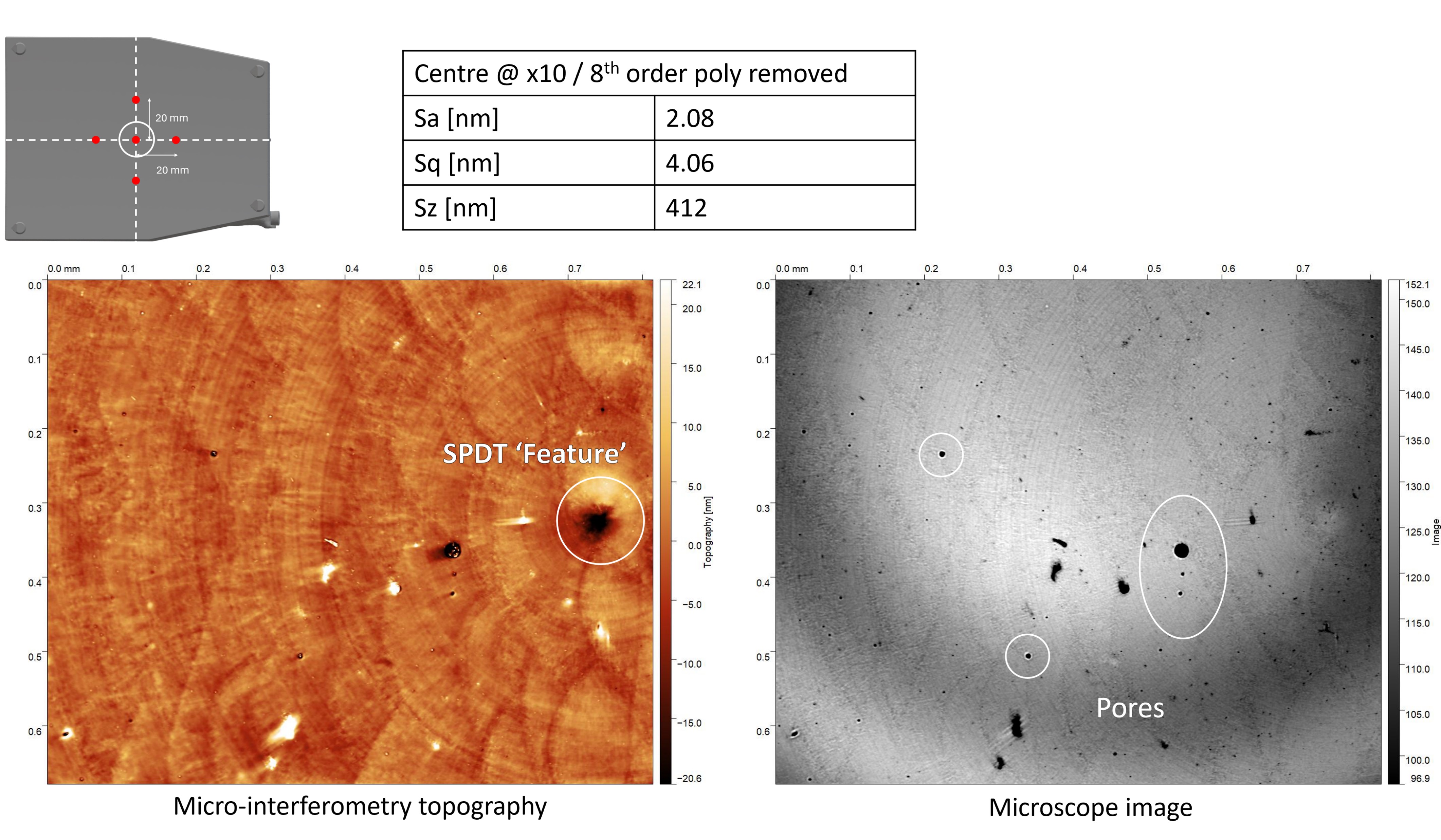}
    \caption{Central measurement site images and roughness values}
    \label{fig:34}
\end{figure}

The surface roughness of the more dense closed-back design was measured at x2.5 and x10 magnification at five different sites on the mirror surface. While the x2.5 data is still being processed, the x10 magnification data highlighted features relating to AM and SPDT processes. During the analysis, an 8\textsuperscript{th}-order polynomial was removed from the data to remove form and waviness and to provide consistency with a parallel project on AM aluminium mirrors. The five different sites can be seen in Figure  \ref{fig:33} along with micro-interferometry topography and microscope images for the four off-centre sites. 
The central site images can be seen in Figure  \ref{fig:34} with the roughness values for that site as well.

Each site had some feature that increased surface roughness, for example, circular cross-sections of pores and the SPDT central point, highlighted in Figure  \ref{fig:34}. The SPDT point is commonly found on SPDT parts and occurs at the centre of rotation when the tool lifts off the part. It was pronounced in this case, as the tool was not calibrated to the nearest micron in the X-axis, however, the emphasis was placed more on the resultant surface roughness and material machinability, than surface form. The pores are another feature that were anticipated on the mirror surface, however there appears to be less than in previous studies\cite{Marcell}.
There are some large spherical pores seen in the images in Figure \ref{fig:33} as well which are most likely keyhole pores which occur when the AM process is too hot. 
Two features adversely affected the polynomial fit within the data: in the (-20, 0) image the white circle, which is not a pore, resulted in a localised form error; and in the image (20,0) the large pore at the edge resulted in an increased form error across the image, as such, the image was cropped before the polynomial subtraction.

\begin{figure}
    \centering
    \includegraphics[width=1\linewidth]{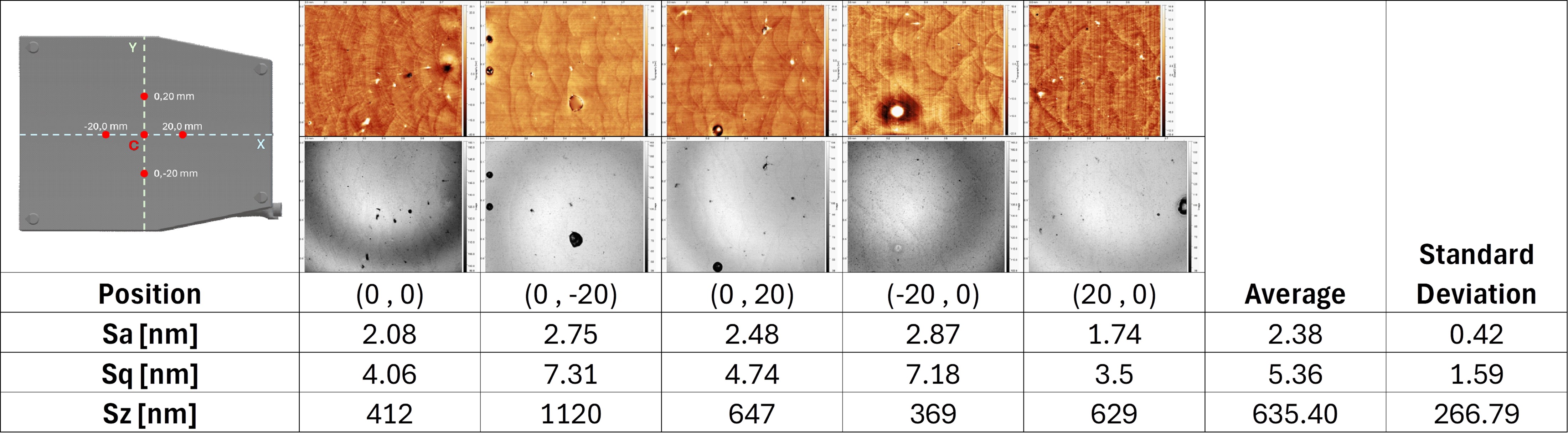}
    \caption{Surface roughness measurements of the more dense, closed-back design}
    \label{table: 4}
\end{figure}

The average roughness (Sa), the root mean square roughness (RMS) (Sq), and the peak to valley (Sz) were measured at each of the five sites, the average for each and the standard deviation can be seen with the rest of the results in Figure \ref{table: 4}.

\section{Conclusion \& Future Work}
\label{sec:8}
The main aim of this project was to re-design and manufacture a nanosat mirror using the advantages of AM, such as light-weighting and part consolidation, aiming to address previously identified challenges in the printing and machining stages, namely print-through and internal stresses.

Three different designs resulted from this process: one open-back design and two closed-back designs. Each design demonstrated the benefits of AM by reducing mass through an internal lattice and topology optimising certain external fixtures. The heavier closed-back design achieved an MR of 36\%, due to a high-density internal lattice. The lighter closed-back design reached an MR of 50\%, using a less dense version of the same lattice, with a thinner strut thickness selected to achieve the 50\% MR. The open-back design had the highest MR at 65\%, achieved by removing the back face of the mirror and infilling it with a sparse TPMS lattice.
Each of the designs used topology optimisation on the parts of the mirror that interfaced with the nanosat optical assembly, reducing mass and also consolidating the 18 parts comprising these features with the main mirror body. This led to a total part consolidation on each design of 19 parts down to 1.

Print-through was investigated for each design; a lattice down-selection process resulted in a Diamond Graph lattice being used on the closed-back designs and a Diamond TPMS lattice being chosen for the open-back design. A cell map that was defined by the surface of the mirror was selected for each design following verification by FEA. 
Three mounting points were required on each design for the SPDT process. These were designed with a down-selected lattice to limit the transfer of internal stresses to the mirror surface. The placement of these mounting points on the back of the mirror were selected to minimise distortion during SPDT.

All three designs were printed successfully and exhibited good detail. The heavier closed-back design was machined first, and the surface roughness was inspected post-SPDT. While some porosity was present on the mirror surface, the average RMS surface roughness was measured at \SI{5.36}{\nm}, which is considered typical of this manufacturing method.

Immediate future work will use X-ray computed tomography to analyse the the sectioned parts ( Figure \ref{fig:30}), assessing the quality in terms of conformance to the CAD and quantifying the presence of porosity. The form error of the heavier closed-back design will be evaluated as well to assess the impact of print-through and mounting stresses on the optical surface. The other two designs will undergo machining and SPDT, having the surface roughness and the form error measured afterwards. The SPDT of these designs will benefit from the experience gained from the first machining and build upon them to gain a better result. For instance, the open-back designs may be machined off-axis in an attempt to eliminate the centre feature and mitigate the scratches on the surface.

\section{Acknowledgements}
The authors acknowledge the UKRI Future Leaders Fellowship ‘Printing the future of space telescopes’ under grant \# MR/T042230/1, as well as the aid of Gregory Campbell, Martin Sanchez and John Horne from CA Models for their expertise and help in the final printing of the parts.

\bibliography{report} 
\bibliographystyle{spiebib} 

\end{document}